\documentclass[aps,preprint,nobibnotes,nofootinbib]{revtex4}
\usepackage{amsfonts}
\usepackage{graphicx}
\usepackage{amsmath}
\usepackage{amssymb}
\usepackage{subfigure}
\usepackage{setspace}
\usepackage{mathrsfs}

\begin{document}

\title{A Rigorous Derivation of Electromagnetic Self-force}

\author{Samuel E. Gralla}
\author{Abraham I. Harte}
\author{Robert M. Wald}
\affiliation{\it Enrico
Fermi Institute and Department of Physics \\ \it University of Chicago
\\ \it 5640 S.~Ellis Avenue, Chicago, IL~60637, USA}

\begin{abstract}
During the past century, there has been considerable discussion and
analysis of the motion of a point charge in an external
electromagnetic field in special relativity, taking into account
``self-force'' effects due to the particle's own electromagnetic
field. We analyze the issue of ``particle motion'' in classical
electromagnetism in a rigorous and systematic way by considering a
one-parameter family of solutions to the coupled Maxwell and matter
equations corresponding to having a body whose charge-current density
$J^a(\lambda)$ and stress-energy tensor $T_{ab} (\lambda)$ scale to
zero size in an asymptotically self-similar manner about a worldline
$\gamma$ as $\lambda \rightarrow 0$. In this limit, the charge, $q$,
and total mass, $m$, of the body go to zero, and $q/m$ goes to a well
defined limit. The Maxwell field $F_{ab}(\lambda)$ is assumed to be
the retarded solution associated with $J^a(\lambda)$ plus a
homogeneous solution (the ``external field'') that varies smoothly
with $\lambda$.  We prove that the worldline $\gamma$ must be a
solution to the Lorentz force equations of motion in the external
field $F_{ab}(\lambda=0)$. We then obtain self-force, dipole forces,
and spin force as first order perturbative corrections to the center
of mass motion of the body. We believe that this is the first rigorous
derivation of the complete first order correction to Lorentz force
motion. We also address the issue of obtaining a self-consistent
perturbative equation of motion associated with our perturbative
result, and argue that the self-force equations of motion that have
previously been written down in conjunction with the ``reduction of
order'' procedure should provide accurate equations of motion for a
sufficiently small charged body with negligible dipole moments and
spin. (There is no corresponding justification for the
non-reduced-order equations.)  We restrict consideration in this paper
to classical electrodynamics in flat spacetime, but there should be no
difficulty in extending our results to the motion of a charged body in
an arbitrary globally hyperbolic curved spacetime.

\end{abstract}

\maketitle

\section{Introduction}\label{sec:intro}

In classical electrodynamics in special relativity,
Maxwell's equations
\begin{align}
\nabla^\nu F_{\mu \nu} & = 4 \pi J_\mu \label{eq:max1}\\
\nabla_{[\mu}F_{\nu\rho ]} & = 0 \label{eq:max2}
\end{align}
are most often considered in the context where the charge current
source $J^\nu$ is a specified function of the spacetime coordinates
$x^\mu = (t,x^i)$. Provided only that $J^\nu$ is conserved,
\begin{equation}
\nabla_\nu J^\nu = 0 \,\, ,
\label{eq:Jcons}
\end{equation}
Maxwell's equations have a well posed initial
value formulation.  Furthermore, since Maxwell's equations are linear
in $F_{\mu \nu}$, it makes perfectly good mathematical sense to allow
$J^\nu$ be a distribution and to seek corresponding distributional
solutions for $F_{\mu \nu}$. In particular, it makes perfectly good
mathematical sense to consider point particle sources, i.e.,
distributional charge-current sources of the form (in global inertial coordinates\footnote{In arbitrary coordinates the right side would contain an additional factor of $(-g)^{-1/2}$ (see eq.\eqref{eq:J1-2} below).})
\begin{equation}
J^\nu = q u^\nu \delta^{(3)}(x^i - z^i(t)) \frac{d\tau}{dt}
\label{eq:ptch}
\end{equation}
where $u^\mu$ is the unit tangent (i.e., 4-velocity) of the worldline
defined by $x^i(t) = z^i(t)$, and $\tau$ is the proper time along this
worldline. Solutions to Maxwell's equations with an arbitrary point particle source of this form exist provided only that this worldline is
timelike and that $q$ is constant along the worldline (as required by
conservation of $J^\nu$, eq.(\ref{eq:Jcons})).

Another problem that is commonly considered in classical
electrodynamics is the motion of a point particle in a given external
field. Here one prescribes a Maxwell field $F^{\textrm{ext}}_{\alpha \beta}$
as a function of spacetime coordinates and one postulates the
Lorentz force equation of motion for the point charge,
\begin{equation}
m a_\alpha = q F^{\textrm{ext}}_{\alpha \beta}u^\beta \,\, ,
\label{eq:lorentz}
\end{equation}
where $a^\alpha \equiv u^\beta \nabla_\beta u^\alpha$ is the
4-acceleration of the worldline.  The Lorentz force equation is
simply a system of second order ordinary differential equations for
determining the worldline to which $u^\alpha$ is the unit
tangent. It is a well posed equation that admits a
unique solution for the motion of the particle
for any specified $F^{\textrm{ext}}_{\alpha \beta}$ (which need not be
a solution to Maxwell's equations) and any specified initial
conditions of the particle.

Since Maxwell's equations are well posed for any specified point
particle source and point particle motion is well posed in any
specified Maxwell field, one might expect that the self-consistent
coupled system, eqs.(\ref{eq:max1})-(\ref{eq:max2}) and (\ref{eq:ptch}), together with
\begin{equation}
m a_\alpha = q F_{\alpha
  \beta}u^\beta \,\, ,
\label{eq:lorentz2}
\end{equation}
would also be well posed.  (Here, eq.(\ref{eq:lorentz2}) differs from
eq.(\ref{eq:lorentz}) in that $F^{\textrm{ext}}_{\alpha \beta}$ has
been replaced by the full Maxwell field $F_{\alpha \beta}$, which
includes the contributions from the ``self-field'' of the point
charge.)  However, it is easily seen that this is not the case: The
full Maxwell field $F_{\alpha \beta}$ is necessarily singular on the
worldline of the particle, and, consequently, eq.(\ref{eq:lorentz2})
is ill-defined. Equations (\ref{eq:max1})-(\ref{eq:max2}) and (\ref{eq:ptch}) together
with eq.(\ref{eq:lorentz2}) simply do not make mathematical sense.

Nevertheless, some authors---most notably Dirac \cite{dirac}---have
attempted to proceed based on formal versions of these equations.
However, in addition to being \textit{ad hoc}, these treatments
necessarily encounter mathematical and physical difficulties, such as
the famous ``infinite mass renormalization''.  In this
picture, the infinite electromagnetic self-energy of a point charge is
compensated by a negatively infinite material mass, rendering the
combined (observable) mass finite. We see no justification for
postulating that material mass could be negative (or infinite).
Rather, as long as matter satisfies positive energy conditions, it is
clear that a true point charge really would have an infinite total
(observable) mass, and is therefore a physically unacceptable object.
In short, the attempt to couple classical matter to electromagnetism
via the point particle hypothesis \eqref{eq:ptch} and the Lorentz
force law \eqref{eq:lorentz} fails on both mathematical and physical
grounds\footnote{Similarly, in general relativity, Einstein's equation
  does not make mathematical sense for point mass sources
  \cite{geroch-traschen}. Physically, matter would collapse to a black
  hole before a point mass limit could be achieved.}.

Thus, it seems clear that the point particle charge-current
\eqref{eq:ptch} and Lorentz force law \eqref{eq:lorentz} cannot be fundamental.  Rather, matter must be modeled as a continuum, with these equations (as well corrections, including self-force effects)
emerging as an approximate description of the motion of ``small bodies''. In fact, much of the earliest work on the self-force problem \cite{abraham, lorentz, schott} effectively followed this approach.  However, a fundamental difficulty arises in this approach.  If one takes a limit to \textit{zero} size in the straightforward way (i.e., at fixed charge and mass), one will simply recover the point particle description and its associated difficulties, such as infinite self-energy.  But if one keeps the body at finite size, then all of the body's internal degrees of freedom will affect its motion, and the description of motion will depend in a complicated way on the details of the structure and initial state of the body.

This latter difficulty is commonly dealt with by postulating a
\textit{rigid} body, thereby eliminating the internal degrees of
freedom. However, the notion of the ``rigidity'' of a body is incompatible with special relativity unless the body follows the orbits of a Killing field. Thus, rigidity places extremely restrictive {\it a priori} constraints on motion in special relativity and is an unacceptable starting point for a general analysis of the motion of a body. A proper continuum model must correctly allow for relativistic
deformations. Mathematically, one should consider only models such that the self-consistent, coupled, Maxwell-charged-matter equations admit a
well-posed initial value formulation, such as a charged fluid or a charged elastic solid (see \cite{geroch} for further discussion).

However, our interest here is in deriving ``universal''
properties of the motion of small bodies that do not depend upon the
details of any particular matter model. Thus, we wish to assume
only the existence of a matter stress-energy
tensor $T^M_{\mu \nu}$, which couples to the electromagnetic field via
conservation of \textit{total} stress-energy,
\begin{equation}
\nabla^\nu \left( T^{M}_{\mu \nu} + T^{EM}_{\mu \nu} \right) = 0 \,\, ,
\label{eq:Tcons}
\end{equation}
where the electromagnetic stress-energy tensor is given by
\begin{equation}
T^{EM}_{\mu \nu} = \frac{1}{4 \pi} \left( F_{\mu \alpha}F_{\nu}^{\
\alpha} - \frac{1}{4}g_{\mu \nu}F_{\alpha \beta}F^{\alpha \beta }
\right) \,\, .
\label{eq:Tem}
\end{equation}
We will refer to the coupled system of
eqs.(\ref{eq:max1}-\ref{eq:Jcons}) and (\ref{eq:Tcons})
as the \textit{Maxwell and matter equations}.  All realistic bodies
composed of continuum matter and charge-currents should satisfy the Maxwell and matter equations, and our results will be derived solely from these equations.\footnote{These equations were also used in \cite{harte2006} as
the starting point in the self-force context.  Very recently, this work was improved \cite{harte2009} in a non-perturbative framework that
provides a complete formalism for describing the motion of extended
bodies, including self-field effects.}  Note that the Maxwell and matter equations are nonlinear in the electromagnetic field.

Since the motion of a
finite-size body will depend on its detailed composition as well as on
the details of
its internal states of motion, we must consider a \textit{limit}
where the body is shrunk down to zero size in order to have a chance
at obtaining a simple, universal description of its motion. As mentioned
above, normally
such a limit is taken at fixed charge and mass, thereby reproducing
the same difficulties as arise if one attempts to work directly
with point particles.
The main new idea in the present work is to consider a mathematically
precise limit---based on suitable one-parameter families of exact solutions
to the Maxwell and matter equations---that will enable us to avoid these
difficulties and thereby
to derive
self-force effects in a mathematically rigorous manner. Following the basic
ideas of
\cite{gralla-wald} (which were formulated in the context of general
relativity), we consider a modified point particle limit, wherein not
only the size of the body goes to zero, but its charge and mass also
go to zero.  More precisely, we will consider a limit where,
asymptotically, only the overall scale of the body changes, so, in
particular, all quantities scale by their naive dimension.  In this
limit, the body itself completely ``disappears'', and its
electromagnetic self-energy goes to zero.  We will show in section
\ref{sec:farzone} that the worldline, $\gamma$, to which the body
shrinks down must be a solution to the Lorentz force equations of
motion. (As we shall also see in section \ref{sec:farzone}, the
electromagnetic self-energy of the body makes a finite, non-zero
contribution to the mass, $m$, that appears in the Lorentz force
equation.)  Self-force effects (as well as dipole force and spin force
effects) then arise as a first order perturbative correction to the
center of mass motion of the body.
We emphasize that
our approach removes from any fundamental status both the point charge
source \eqref{eq:ptch} and the Lorentz force law \eqref{eq:lorentz}.
Rather, the notion of a ``point charge'' and the Lorentz force
equation that it satisfies will be \textit{derived} as a lowest order
description of the electromagnetic field and bulk motion of a small
(but extended) body\footnote{Of course, the body of most interest to
  experimental phenomena, the electron, is an intrinsically
  quantum-mechanical object. Our results will apply to electrons only
  to the extent that a classical description of the motion of an
  electron can be justified.} whose exact evolution is governed by
the Maxwell and matter equations (\ref{eq:max1}-\ref{eq:Jcons}),
and \eqref{eq:Tcons}.

Our derivation will enable us greatly
clarify what is perhaps the most famous difficulty of the
electromagnetic self-force problem: ``runaway solutions''.  The output
of the classic Abraham \cite{abraham}, Lorentz \cite{lorentz} and
Dirac \cite{dirac} analyses (as well as of many other analyses) is the
following equation:
\begin{equation}\label{eq:ALD-intro}
m a_\alpha = q F^{\textrm{ext}}_{\alpha \beta} u^\beta +
\frac{2}{3}q^2\left( g_{\alpha \beta} + u_\alpha u_\beta
\right) \dot{a}^\beta \,\, ,
\end{equation}
usually known as the Abraham-Lorentz-Dirac (ALD) equation.  The second
term on the right side of this equation corresponds to the
``self-force'' on the charge resulting from its own electromagnetic
field.  This term makes the ALD equation \textit{third}-order
in time, rendering the status of its initial value formulation---and
consequently its role in predicting actual particle motion---at best
unclear.  Furthermore, eq.~(\ref{eq:ALD-intro}) has solutions whose
acceleration grows exponentially in time---behavior that is not
observed in nature.  Clearly, eq.~(\ref{eq:ALD-intro}) is
disastrously pathological as it stands.

Historically, there have been three basic responses to these problems.
The first response views eq.~(\ref{eq:ALD-intro}) to be a genuine
prediction of classical particle electrodynamics, and seeks a
resolution of the runaway difficulties by a treatment of the problem
within the framework of quantum mechanics or quantum electrodynamics
\cite{QM}. If so, then classical electrodynamics would be seriously
deficient.  The second response \cite{dirac} holds that
eq.~(\ref{eq:ALD-intro}) may be salvaged by introducing a set of rules
designed to eliminate runaway behavior \cite{dirac}.  However, such
rules are necessarily acausal in nature and would appear to be
incompatible with the underlying causal behavior of classical
electrodynamics. The third response maintains that the equation should
be modified by the ``reduction of order'' procedure
\cite{reduction-of-order}.  This procedure is based on the observation
that eq.~\eqref{eq:ALD-intro} is only expected to be valid to
order $q^2$.  Therefore, if we replace the $\dot{a}^\beta$ term in
eq.~\eqref{eq:ALD-intro} by the value it would take for Lorentz force
motion in the background field $F^{\textrm{ext}}_{\alpha \beta}$, we
should get an equation that is ``equally valid'' to order $q^2$.
Thus, to order $q^2$, we may replace $a^\alpha$ on the right side of
eq.~(\ref{eq:ALD-intro}) by
\begin{equation}
a_\alpha = \frac{q}{m} F^{\textrm{ext}}_{\alpha \beta} u^\beta
\label{replacea}
\end{equation}
and we may replace $\dot{a}^\alpha$ by
\begin{eqnarray}
\dot{a}_\beta &=& u^\gamma \nabla_\gamma \left( \frac{q}{m} F^{\textrm{ext}}_{\beta \sigma}
u^\sigma \right)  \nonumber \\
&=& u^\sigma u^\gamma \nabla_\gamma \left( \frac{q}{m} F^{\textrm{ext}}_{\beta \sigma}
 \right) + \left( \frac{q}{m} F^{\textrm{ext}}_{\beta \sigma}
 \right) \frac{q}{m} F^{\textrm{ext} \sigma}_{\quad \ \ \delta} u^\delta
\label{replaceadot}
\end{eqnarray}
The resulting equation
\begin{equation}\label{eq:ALD-reduced-intro}
m a_\alpha = q F^{\textrm{ext}}_{\alpha \beta} u^\beta +
\frac{2}{3}q^2\left( g_\alpha^{\ \beta} + u_\alpha u^\beta
\right) \left[ u^\sigma u^\gamma \nabla_\gamma \left( \frac{q}{m} F^{\textrm{ext}}_{\beta \sigma}
 \right) + \left( \frac{q}{m} F^{\textrm{ext}}_{\beta \sigma}
 \right) \frac{q}{m} F^{\textrm{ext} \sigma}_{\quad \ \ \delta} u^\delta \right]
\end{equation}
is known as the ``reduced order'' Abraham-Lorentz equation. This
equation is free from the ``runaway'' solutions that plague
eq.\eqref{eq:ALD-intro}, and has a standard, second-order initial
value formulation. Thus, eq.~(\ref{eq:ALD-reduced-intro}) appears to
be satisfactory on both mathematical and physical grounds. However,
the reduction of order procedure has the appearance of being rather
{\it ad hoc}, and if eq.~(\ref{eq:ALD-intro}) is, in some sense, more
fundamentally correct, it is difficult to justify replacing this
equation with eq.~(\ref{eq:ALD-reduced-intro}) other than on purely
pragmatic grounds.

Our analysis sheds considerable light on this issue.  We will prove in
section \ref{sec:nearzone} that the first order deviation from Lorentz
force motion is given by\footnote{We also obtain equations for the
  time evolution of $S_{ab}$ and $\delta m$ (see
  eqs. \eqref{eq:spin-cov} and \eqref{eq:mass-cov} below).  However,
  the electromagnetic dipole $Q^{ab}$ does not have any associated
  conservation law and can be ``changed at will by the body itself'',
  so there exists no universal evolution equation for this quantity.
  Similar remarks should hold for all higher moments (of mass and of
  charge), although these moments would show up only at higher orders
  in our perturbation expansion.}
\begin{equation}
\label{eq:EOM-intro}
\delta [\hat{m} a_{\alpha} ] = \delta [ q F^{\textrm{ext}}_{\alpha \beta} u^{\beta} ] + \left( g_{\alpha}^{ \ \beta} + u_{\alpha} u^{\beta} \right) \bigg \{ \frac{2}{3} q^2 \frac{D}{d\tau}a_{\beta} - \frac{1}{2} Q^{\gamma \delta}\nabla_{\beta} F^{\textrm{ext}}_{\gamma \delta} + \frac{D}{d\tau} \left( a^{\gamma} S_{\gamma \beta} - 2 u^{\delta} Q^{\gamma}_{\ [\beta} F^{\textrm{ext}}_{\delta]\gamma} \right) \bigg \}.
\end{equation}
In this equation $\delta[\dots]$ refers to the first perturbative
correction of $\dots$, whereas all other quantities are lowest
non-vanishing order.  The parameters $q$, $\hat{m}$, $Q^{\alpha \beta}$, and
$S^{\alpha \beta}$ are, respectively, the charge, mass,\footnote{The perturbed
  mass, $\delta \hat{m}$, includes a contribution from the interaction
  energy of the electric dipole with the external field (see
  eq.(\ref{deltamhat}) below).} electromagnetic dipole moment, and
spin of the body. Here the electromagnetic dipole $Q^{\alpha \beta}$ can be
decomposed in terms of the electric dipole moment $p^{\alpha}$ and magnetic dipole
moment $\mu^{\alpha}$ by
\begin{equation}\label{eq:DipoleDecomposition}
Q^{\alpha \beta} = 2 u^{[\alpha} p^{\beta]} - \epsilon^{\alpha \beta \gamma \delta}
u_\gamma \mu_\delta
\end{equation}
with $u^{\alpha} p_{\alpha} = u^{\alpha} \mu_{\alpha} = 0$, and the spin tensor $S^{\alpha \beta}$ can be expressed
in terms of the spin vector $S^{\alpha}$ by
\begin{equation}\label{eq:SpinVectToTens}
S^{\alpha \beta} = \epsilon^{\alpha \beta \gamma \delta} u_{\gamma} S_{\delta} \,\, .
\end{equation}

The first term on the right side of equation \eqref{eq:EOM-intro}
is the first order correction to the ordinary Lorentz force law. The
first term in braces corresponds to the ALD self-force. The second
term in braces is a relativistic form of the usual dipole forces familiar from electrostatics and magnetostatics. The remaining terms are additional dipole and spin forces relevant in nonstationary situations.  A non-relativistic version of these force terms is given in eq.\eqref{eq:non-rel} below.

In the case of a body with negligible electric and magnetic
dipole moments and negligible spin, only the first two terms of
equation \eqref{eq:EOM-intro} remain.  In this case, we
see that the form of the ALD equation (\ref{eq:ALD-intro}) is
recovered as the leading order perturbative correction to Lorentz
force motion for a sufficiently small charged body.  However, since the
acceleration $a^{\mu}$ present in the ALD term refers to the
``background motion'' (i.e., a solution to the Lorentz force equations
of motion in the background electromagnetic field), \textit{no runaway
behavior occurs} in the context of perturbation theory.

Although \eqref{eq:EOM-intro} has been rigorously derived as the
leading order perturbative correction to Lorentz force motion, small
deviations from Lorentz force motion can accumulate over
time. Eventually, the deviation from the zeroth order motion should
become large, so one would not expect that solutions to
\eqref{eq:EOM-intro} would provide an accurate description of motion
at late times. However, as argued in \cite{gralla-wald}, if the
deviations from Lorentz force motion are \textit{locally} small, it
may be possible to find a ``self-consistent perturbative equation''
based upon eqs.(\ref{eq:lorentz}) and \eqref{eq:EOM-intro} that
provides a good, global-in-time description of the motion. We shall
argue in section \ref{sec:self-consistent} that in the case of a body with negligible electric and magnetic dipole moments and negligible spin, the reduced order Abraham-Lorentz equation (\ref{eq:ALD-reduced-intro}) provides such a desired self-consistent perturbative equation. This provides a solid justification for using eq.(\ref{eq:ALD-reduced-intro}) to describe the motion of sufficiently small charged bodies (with negligible dipole moments and spin). There is no corresponding justification for using the ALD equation (\ref{eq:ALD-intro}) to describe the motion of charged bodies.

Our notation and conventions are as follows.  Greek indices $\mu,\nu,..$ will refer to spacetime coordinate components, whereas mid-alphabet Latin indices $i,j,...$ will refer to spatial coordinates only.  A ``$0$'' will denote the time component.  Early-alphabet Latin indices $a,b,...$ will refer to abstract spacetime indices \cite{wald}.  (Typically, $a,b,...$ will be used in diffeomorphism covariant expressions, whereas $\mu,\nu,...$ will be used for expressions that take a special form in the coordinates being used.)  Our (flat) metric $g_{ab}$ has signature $(-1,1,1,1)$.  We normalize the Levi-Civita symbol $\epsilon_{abcd}$ so that $\epsilon_{0123}=+1$.

\section{Assumptions}\label{sec:assumptions}

We consider a one-parameter family of charged bodies described by a
charge-current vector $J^\mu(\lambda)$ and a matter stress-energy tensor
$T^{M}_{\mu \nu}(\lambda)$. The electromagnetic field $F_{\mu
  \nu}(\lambda)$ is assumed to satisfy Maxwell's equations
(\ref{eq:max1}) and (\ref{eq:max2}), the charge-current is assumed to
be conserved, eq.(\ref{eq:Jcons}), and the total stress energy tensor is
assumed to be conserved, eq.(\ref{eq:Tcons}).  Since we seek relations that hold universally for all bodies, we specify no additional ``constitutive'' relations, and impose only
eqs.(\ref{eq:max1})-(\ref{eq:Jcons}) and (\ref{eq:Tcons}) in our analysis.

As discussed in the previous section, we wish to consider a limit
where the body shrinks down to zero size, but in such a way that the
mathematical (and physical) inconsistencies associated with point particles are
avoided. Following the basic ideas of \cite{gralla-wald}, this can be
done by considering a limit where not only the size of the body goes
to zero, but its charge and mass also go to zero in such a way that,
asymptotically, the body continues to ``look the same'' except for its
overall scale. In \cite{gralla-wald}, this idea was implemented in a
relatively indirect way by postulating the existence of a ``scaled
limit'' of the exterior field of the body together with the existence
of an ``ordinary limit'' and a ``uniformity condition'' on these
limits. In the context of general relativity, it was necessary in
\cite{gralla-wald} to formulate the conditions on the one-parameter
family in terms of the exterior field because we wished to consider
strong field objects (such as black holes), which do not admit a
simple description in terms of a stress-energy source. However, in the
present context of classical electrodynamics in Minkowski spacetime,
all of the relevant information about the body and the
electromagnetic field is contained in the specification of the charge
current density $J^\mu$, the matter stress-energy tensor $T^{M}_{\mu
  \nu}$, and ``external'' electromagnetic field, i.e., the homogeneous
solution of Maxwell's equations that describes the difference between
the actual electromagnetic field and the retarded solution associated
with $J^\mu$. Consequently, we will formulate our conditions in terms
of these quantities.

We wish to consider a one-parameter family $\{F_{\mu
  \nu}(\lambda),J^\mu(\lambda),T^{M}_{\mu \nu}(\lambda)\}$ of
solutions to eqs.(\ref{eq:max1})-(\ref{eq:Jcons}) and (\ref{eq:Tcons})
having the property that the worldtube, ${\mathcal W}(\lambda)$,
containing the supports of $J^\mu(\lambda)$ and $T^{M}_{\mu
  \nu}(\lambda)$ shrinks down to a timelike worldline $\gamma$ as
$\lambda \rightarrow 0$. Furthermore, we wish to impose the condition
that this ``shrinking down'' asymptotically corresponds merely to a
``change of scale''. To get a feeling for what this entails, fix a
time, $t_0$, on $\gamma$, and consider the charge density,
$\rho(\lambda, x^i)$, on the hyperplane orthogonal to $\gamma$ at
$t_0$ given by
\begin{equation}
\rho(\lambda, x^i) = \lambda^{-2} \tilde{\rho}(\frac{x^i - z^i}{\lambda})
\label{eq:exscale}
\end{equation}
where $\tilde{\rho}$ is a smooth function on ${\bf R}^3$ that vanishes
for $r>R$ for some $R>0$.  Then $\rho(\lambda, x^i)$ vanishes for
$r>\lambda R$, i.e., the support of $\rho$ shrinks to $z^i$ as
$\lambda \rightarrow 0$. It also is clear that for the one-parameter
family, eq.(\ref{eq:exscale}), the charge density ``retains its shape''
exactly as it shrinks to the origin. Finally, the overall scaling of
$\lambda^{-2}$ in eq.(\ref{eq:exscale}) ensures that the total charge
also goes to zero proportionally to $\lambda$. Thus,
eq.(\ref{eq:exscale}) describes the kind of ``scaling to zero size'' that
we seek.

However, since a body undergoing non-uniform motion cannot remain
``rigid'', we must allow for time dependence in the manner in which
the charge-current and body stress-energy scale to zero size,
particularly in view of the fact that the body may find itself in a
different external field at different times. In addition, the exact
scaling represented by eq.(\ref{eq:exscale}) is too strong an assumption;
in particular, it could hold, at most, for a particular choice of time
slicing and a particular choice of spatial coordinates on these
slices. Furthermore, although we could always impose
eq.(\ref{eq:exscale}) exactly as an initial condition at one time, there
is no reason to believe that dynamical evolution would preserve this
exact scaling at later times. Instead, it would be much more
reasonable to require that the type of scaling represented by
eq.(\ref{eq:exscale}) hold only asymptotically as $\lambda \rightarrow
0$.

In order to formulate our conditions on $J^\mu(\lambda)$ and $T^{M}_{\mu
  \nu}(\lambda)$, it is useful to adopt a specific choice of coordinate system, namely Fermi normal coordinates (see, e.g., \cite{poisson}).  These coordinates $(T,X^i)$ are defined with reference to a timelike worldline, $\gamma$.  The surfaces of constant $T$ are orthogonal to $\gamma$, and the $X^i$ measure spatial distance within those surfaces.  With this coordinate choice we may state our conditions on $J^\mu(\lambda)$ and $T^{M}_{\mu \nu}(\lambda)$ as follows:
\begin{itemize}
\item
(i) We require that there be a smooth, timelike worldline $\gamma$ such that in Fermi normal coordinates, $(T,X^i)$, based on $\gamma$, we have
  $J^\mu(\lambda,T,X^i) = \lambda^{-2}
  \tilde{J}^\mu(\lambda,T,X^i/\lambda)$ and $T^{M}_{\mu
    \nu}(\lambda,T,X^i) = \lambda^{-2} \tilde{T}^{M}_{\mu
    \nu}(\lambda,T,X^i/\lambda)$, where $\tilde{T}^{M}$ and
  $\tilde{J}$ are smooth functions of their arguments. Furthermore,
  the supports of $\tilde{T}^{M}$ and $\tilde{J}$ in the last
  variables---i.e., the slots into which $X^i/\lambda$ have been
  inserted---are contained within a radius $R$, so that the
  $J^\mu(\lambda)$ and $T^{M}_{\mu \nu}(\lambda)$ are nonvanishing
  only within a worldtube, ${\mathcal W}(\lambda)$, of radius $\lambda
  R$ about $\gamma$.
\end{itemize}
Note that the smooth dependence on both $T$ and $\lambda$ permitted for
$\tilde{J}^\mu$ and $\tilde{T}^{B}_{\mu \nu}$ in this condition
addresses the concerns of the previous paragraph.

We now present these assumptions in general coordinates.  Let $(t,x^i)$ denote any coordinates that are smoothly related to Fermi normal coordinates in the neighborhood of $\gamma$ in which the Fermi normal coordinates are defined and are such that $\nabla_\mu t$ is timelike. In these new coordinates, the worldline $\gamma$ will be given by an equation of the form $x^i = z^i(t)$. Since the Fermi coordinate
$X^i$ is a smooth function that vanishes on $\gamma$, we must have
\begin{equation}
X^i = \sum_j [x^j - z^j(t)] h_{ij}(t,x^k) \,\, ,
\end{equation}
where $h_{ij}$ is smooth.
Writing $T = T(t, x^k)$, writing
\begin{equation}
x^k = \lambda\frac{x^k - z^k(t)}{\lambda} + z^k(t) \,\, .
\end{equation}
and substituting into the formula for the charge current density in
the new coordinates, we find that the components of the charge current
density in the new coordinates take the form
\begin{equation}
J^{\mu}(\lambda,t,x^i) = \lambda^{-2}
\tilde{J'}^{\mu}(\lambda,t,[x^i-z^i(t)]/\lambda) \,\, ,
\label{eq:Jnew}
\end{equation}
where $\tilde{J'}^{\mu}$ is a smooth function of its arguments.
Furthermore, since $\nabla_\mu t$ is timelike, each surface of
constant $t$ must intersect ${\mathcal W}(\lambda)$ in a compact set,
which implies that $\tilde{J'}^{\mu}$ is of compact support in the
last variables. Note that, in particular, the form eq.(\ref{eq:Jnew}) holds in
global inertial coordinates. By reversing the argument, it is easily
seen that assumption (i) on $J^\mu(\lambda)$ as stated in Fermi normal
coordinates is equivalent to eq.(\ref{eq:Jnew}) holding in some
coordinates with $\nabla_\mu t$ timelike, together with the
requirement that the support of $J^\mu(\lambda)$ be contained in a
worldtube of proper distance $\lambda R$ from $\gamma$.  Similar
results hold, of course, for $T^{M}_{\mu \nu}(\lambda,t,x^i)$.

Our assumption about the electromagnetic field $F_{\mu \nu}(\lambda)$
is as follows:
\begin{itemize}
\item
(ii) We have $F_{\mu \nu} = F_{\mu \nu}^{\textrm{ext}} + F_{\mu
  \nu}^{\textrm{self}}$, where $F_{\mu \nu}^{\textrm{self}}$ is the
  retarded\footnote{It may appear that we are introducing a time
    asymmetry in our assumptions by choosing $F_{\mu
      \nu}^{\textrm{self}}$ to be the retarded solution.  However,
    choosing $F_{\mu \nu}^{\textrm{self}}$ to be the advanced solution
    results in an equivalent set of assumptions, since the
    retarded minus advanced solution is a homogeneous solution that
    can be shown to be smooth in $(\lambda,t,x^i)$.} solution of Maxwell's
  equations with source $J^\mu(\lambda)$ and $F_{\mu
    \nu}^{\textrm{ext}}$ is a homogeneous solution of Maxwell's
  equation that is jointly smooth function of $\lambda$ and the
  spacetime point.
 \end{itemize}

Assumptions (i)-(ii) about our one-parameter family together with
eqs.(\ref{eq:max1})-(\ref{eq:Jcons}) and (\ref{eq:Tcons})
constitute the entirety of what we shall assume in this paper.

In Appendix \ref{sec:app-scaling}, we show that in arbitrary smooth coordinates
$(t,x^i)$, the retarded solution, $F_{\mu
  \nu}^{\textrm{self}}(\lambda)$, with source of the form
eq.(\ref{eq:Jnew}) takes the form
\begin{equation}
F_{\mu \nu}^{\textrm{self}}(\lambda,t,x^i) = \lambda^{-1}
\tilde{F}_{\mu \nu}(\lambda,t,[x^i-z^i(t)]/\lambda) \,\, ,
\label{eq:Fform}
\end{equation}
where $\tilde{F}$ is a smooth function of its arguments.

Finally, we relate the assumptions about our one-parameter family made
here to the assumptions made in \cite{gralla-wald}. As previously
mentioned, the assumptions of \cite{gralla-wald} were formulated
entirely in terms of the behavior of the ``exterior field'' of the
body, i.e., the behavior of the spacetime metric outside of the
body. Assumption (i) of \cite{gralla-wald} (``existence of the
ordinary limit'') corresponds here to the requirement that $F_{\mu
\nu}$ be jointly smooth in $(\lambda,t,x^i)$ except on $\gamma$ and
that $F_{\mu \nu}(\lambda=0)$ be smoothly extendible to $\gamma$. That
this condition holds follows immediately from our assumption (ii)
together with the fact that $F_{\mu \nu}^{\textrm{self}}$ goes to zero
at fixed $(t,x^i)$ with $x^i \neq z^i(t)$ as $\lambda \rightarrow
0$. The analog of assumption (ii) of \cite{gralla-wald} (``existence
of the scaled limit'') follows immediately from assumption (ii) and
eq.(\ref{eq:Fform}). We will consider this scaled limit in detail in section \ref{sec:nearzone} below. Finally, the analog of the ``uniformity condition'' (iii) of
\cite{gralla-wald} is as follows: Define $\alpha = r = \sqrt{\sum [x^i
    - z^i(t)]^2}$ and define $\beta=\lambda/r$. Write $F_{\mu \nu}$ as a
function of $(\alpha, \beta, t, \theta, \phi)$. The desired condition
is that at fixed $(t, \theta, \phi)$, $\lambda F_{\mu \nu}$ is jointly
smooth in $(\alpha, \beta)$ at $(0,0)$. Since $\lambda F_{\mu
  \nu}^{\textrm{ext}}$ is easily seen to satisfy this condition, the
analog of the uniformity condition of \cite{gralla-wald} will hold
here provided that $\tilde{F}_{\mu \nu}$ is smooth in $(\alpha,
\beta)$ at
$(0,0)$ at fixed $(t, \theta, \phi)$. In appendix \ref{sec:app-scaling}, we
show that not only is $\tilde{F}_{\mu \nu}$ smooth in $(\alpha,
\beta)$, but is of the form of $\beta^2$ times a smooth function of
$(\alpha, \beta)$.

Thus, the assumptions we have made above imply analogs of the
assumptions made in \cite{gralla-wald}. Since our assumptions here are
quite simple and straightforward as compared with the assumptions made
in \cite{gralla-wald}, this may be viewed as providing justification
for the assumptions made in \cite{gralla-wald}. Since the assumptions (i)-(iii) of
 \cite{gralla-wald} are essentially what is needed to justify the use of
``matched asymptotic expansions''  \cite{matched-expansions,poisson},
we have thus effectively also provided justification for the use matched asymptotic expansions.

\section{The Far-Zone Limit and Lorentz Force Motion}
\label{sec:farzone}

As in \cite{gralla-wald}, as $\lambda \rightarrow 0$ it will be useful
to consider both a ``far-zone'' limit (wherein $x^\mu$ is held fixed)
and a ``near zone'' limit (wherein $\bar{x}^\mu \equiv ([t-t_0]/\lambda,
[x^i - z^i(t_0)]/\lambda)$ is held fixed). Roughly speaking, the ``far
zone'' limit corresponds to how the body and the electromagnetic field
appear to an observer at a fixed radius from $\gamma$ as the body
shrinks down to $\gamma$. The near zone limit describes the
appearance to an observer who ``follows the body in'' toward $\gamma$ at time $t_0$---and correspondingly rescales units---as the body shrinks. We use the terminology ``far zone'' and ``near
zone'' because of the close correspondence to the way these terms are
used in matched asymptotic expansion analyses.  However, we will not need to do
any ``matching'' in our analysis, and any of our calculations
can be done in either the ``far'' or ``near'' zone pictures.  In this section, we treat the far-zone limit.  We will show that at first order in $\lambda$, the far zone description of the body is that of a point particle, and that the
worldline $\gamma$ satisfies the Lorentz force equation of motion
(\ref{eq:lorentz}). Interestingly, we will also see that the
electromagnetic self-energy of the body makes a non-zero, finite
contribution to the particle's mass.

It is clear from assumption (i) of the previous section that
$J^\mu(\lambda)$ and $T^{M}_{\mu \nu}(\lambda)$ as well as all of
their $\lambda$-derivatives go to zero pointwise as $\lambda
\rightarrow 0$ and any fixed $x^\mu \notin \gamma$. On the other hand,
$J^\mu(\lambda)$ and $T^{M}_{\mu \nu}(\lambda)$ and their
$\lambda$-derivatives do not approach any limit at all for
$x^\mu \in \gamma$. Nevertheless, $J^\mu(\lambda)$, $T^{M}_{\mu
\nu}(\lambda)$ and their $\lambda$-derivatives have well defined
limits as distributions as $\lambda \rightarrow 0$. To see this, we
recall that associated to any locally $L^1$ function $H(t,x^i)$ on
spacetime is the distribution, $D_H$, defined by the following action on
a smooth test function $f(t,x^i)$ of compact support,
\begin{equation}
D_H[f] = \langle H,f \rangle = \int H(t,x^i) f(t,x^i) \sqrt{-g} d^4 x \,\, .
\end{equation}
Thus, viewed as a distribution, the charge-current
$J^\mu(\lambda)$ is given by
\begin{align}\label{eq:manip}
D_J(\lambda)[f] & = \int J^\mu(\lambda,t,x^i)
f_\mu(t,x^i) \sqrt{-g} \ dt d^3x \nonumber \\ & = \int \lambda^{-2}
\tilde{J}^\mu(\lambda, t, [x^i-z^i(t)]/\lambda) f_\mu(t,x^i) \sqrt{-g}
\ dt d^3x \nonumber \\ & = \lambda \int
\tilde{J}^\mu(\lambda,t,\bar{x}^i) f_\mu(t,z^i(t) + \lambda \bar{x}^i)
\sqrt{-g} \ dt d^3\bar{x}   \nonumber \\ & = \lambda \int dt \frac{d\tau}{dt}
f_\mu(t,z^i(t)) \int d^3\bar{x} \sqrt{g_3} \tilde{J}^\mu(0,t,\bar{x}^i)
+ O(\lambda^2) \,\, .
\end{align}
(Here we work in arbitrary smooth coordinates $(t,x^i)$ and have used
eq.(\ref{eq:Jnew}) in the second line; we also have dropped the prime on $\tilde{J}^\mu$.)
Thus, we see that, viewing $J^\mu(\lambda)$
as a one-parameter family of distributions, we have
\begin{align}
J^{(0)\mu} & \equiv \lim_{\lambda \rightarrow 0} J^\mu(\lambda)
= 0 \label{eq:J0} \\
J^{(1)\mu} &  \equiv \lim_{\lambda \rightarrow 0} \frac{\partial}
{\partial \lambda} J^\mu(\lambda)
= \mathcal{J}^\mu(t) \frac{\delta(x^i-z^i(t))}{\sqrt{-g}} \frac{d\tau}{dt} ,\label{eq:J1}
\end{align}
with
\begin{equation}
\mathcal{J}^\mu(t) \equiv \int \tilde{J}^\mu(\lambda=0,t,\bar{x}^i) \sqrt{g_3}\ d^3\bar{x},
\end{equation}
Higher-order derivatives of $J^\mu$ with respect to $\lambda$ will
similarly have the form of distributions with support on
$\gamma$.  For example, $J^{(2)\mu}$ will contain a delta function, which will contribute a correction to the charge of the body, along with a derivative of delta function, which will provide the leading order contribution to the electromagnetic dipole moment.  In general, the $n$th order charge-current $J^{(n)\mu}$ will have the form of the sum of derivatives of the delta function up to order $n$, where the coefficient of the $m$th order delta function is the $(n-m)$th correction to the $(n-1)$st moment.

We can obtain considerably more information about the form of
$J^{(1) \mu}$, eq.(\ref{eq:J1}), by making use of conservation of
$J^\mu$, eq.(\ref{eq:Jcons}).  For any test function $f$, we have
\begin{equation}
\int \nabla_\nu J^\nu(\lambda) f \sqrt{-g} d^4 x = 0 \,\, ,
\label{eq:Jcons-dist}
\end{equation}
for all $\lambda > 0$.  Integrating by parts, differentiating with
respect to $\lambda$, and taking the limit as $\lambda \rightarrow 0$,
we obtain
\begin{equation}\label{eq:Jcons-dist2}
\int \mathcal{J}^\mu(t) [\nabla_\mu f](t,z^i(t)) d\tau = 0,
\end{equation}
for all test-functions $f$.  Now, for any $i=1,2,3$, choose $f$ to be
a test-function of the form $f = X^i c(t) h(X^j)$, where $X^i$ is a
Fermi normal spatial coordinate and $c$ and $h$ are of compact support
with $h(X^j)=1$ in a neighborhood of the origin.  Equation
\eqref{eq:Jcons-dist2} then yields
\begin{equation}
\int \mathcal{J}^\mu (t) [\nabla_\mu X^i] c(t) d\tau = 0,
\end{equation}
which implies that the projection of $\mathcal{J}^\mu(t)$ orthogonal
to $\gamma$ vanishes for all $t$, i.e., $\mathcal{J}^\mu$ is of the
form $\mathcal{J}(t) u^\mu$.  Using this result in equation
\eqref{eq:Jcons-dist2} and integrating by parts, we find that
$\dot{\mathcal{J}}(t)=0$.  We thus obtain
\begin{equation}
J^{(1)\mu} = q u^\mu \frac{\delta(x^i-z^i(t))}{\sqrt{-g}}\frac{d\tau}{dt},\label{eq:J1-2}
\end{equation}
where the (lowest-order) total \textit{charge} $q$ is given by
\begin{equation}\label{eq:q}
q \equiv \int \tilde{J}^0 (\lambda=0,t,\bar{x}^i) \sqrt{g_3} \ d^3\bar{x}
\end{equation}
and is independent of time $t$.  Thus, to first order in $\lambda$, the charge current of the body is precisely the standard form of a structureless point charge moving on the worldline $\gamma$.

The electromagnetic field $F_{\mu \nu}$ is the sum of
$F^{\textrm{ext}}_{\mu \nu}$ and $F^{\textrm{self}}_{\mu \nu}$.  The
external field $F^{\textrm{ext}}_{\mu \nu}$ is smooth in $\lambda$ and
satisfies the homogeneous Maxwell's equations at all
$\lambda$. Therefore, $F^{\textrm{ext}}_{\mu \nu} (\lambda)$ has a
straightforward perturbation expansion in $\lambda$, and all
$\lambda$-derivatives of $F^{\textrm{ext}}_{\mu \nu}$ satisfy the
homogeneous Maxwell's equations at each order in perturbation theory.

From eq.~\eqref{eq:Fself-ab} of appendix \ref{sec:app-scaling}, we see
that the self-field is of the form
\begin{equation}
  F_{\mu \nu}^{\mathrm{self}} = \frac{\lambda}{r^2} \mathcal{F}_{\mu \nu}(t, r, \lambda/r, \theta, \phi) \,\, ,
\end{equation}
where, for any fixed ($t,\theta,\phi$), $\mathcal{F}_{\mu \nu}$ is
smooth near 0 in its second and third arguments.  Using this property,
we may Taylor expand $F_{\mu \nu}^{\mathrm{self}}$ to any finite orders $N,M$,
\begin{align}
F_{\mu \nu}^{\mathrm{self}}(t,r,\theta,\phi) & = \frac{\lambda}{r^2}
\sum_{n=0}^N \sum_{m=0}^M r^n \left(\frac{\lambda}{r}\right)^m
(\mathcal{F}_{\mu \nu})_{nm}(t,\theta,\phi) + O\left(r^{N+1}\right) +
O\left((\lambda/r)^{M+1}\right) \nonumber \\ &= \sum_{n=0}^N
\sum_{m=0}^M \lambda^{m+1} r^{n-m-2}(\mathcal{F}_{\mu
  \nu})_{nm}(t,\theta,\phi) + O\left(r^{N+1}\right) +
O\left((\lambda/r)^{M+1}\right) \label{eq:F-series}
\end{align}
This gives the general form of the far-zone expansion of
$F^{\textrm{self}}_{\mu \nu}$ near $r=0$.

We can obtain explicit forms for the terms arising in the far-zone
expansion of $F^{\textrm{self}}_{\mu \nu}$ by using the fact that for all $\lambda$,
$F^{\textrm{self}}_{\mu \nu}(\lambda)$ is the
retarded solution associated with the charge-current $J^\mu(\lambda)$, so
the $n$-order perturbative self-field $F^{\textrm{self},(n)}_{\mu
  \nu}$ is the retarded solution associated with the distributional
source $J^{(n)\mu}$.  In particular, the zeroth order self-field
vanishes
\begin{equation}
F^{\textrm{self},(0)}_{\mu \nu} = 0
\end{equation}
and the first order self-field $F^{\textrm{self},(1)}_{\mu \nu}$ is just
the standard Lienard-Wiechert solution associated with the point
charge source, eq.(\ref{eq:J1}). In Fermi normal coordinates about
$\gamma$, $F^{\textrm{self},(1)}_{ \mu \nu}$ takes the explicit form
\begin{align}
\begin{split}
F^{\textrm{self},(1)}_{i0} & = q \bigg \{ \frac{n_i}{r^2} + \frac{1}{r}
\left( - \frac{1}{2}(a_j n^j)n_i - \frac{1}{2} a_i \right) + \\ &
\qquad + \frac{3}{8} (a_j n^j)^2 n_i + \frac{3}{4} (a_j n^j) a_i +
\frac{1}{8} a_j a^j n_i + \frac{1}{2} \dot{a}_0 n_i + \frac{2}{3}
\dot{a}_i + O(r) \bigg \} \nonumber
\end{split} \\
F^{\textrm{self},(1)}_{ij} & = - q \frac{1}{2}\dot{a}_{[i}n_{j]} + O(r) \,\, ,
\label{eq:F1}
\end{align}
where $n^i=x^i/r$.

We now turn our attention to the stress-energy tensor. By assumption
(i), the matter stress-energy $T^M_{\mu \nu} (\lambda)$ has the same type of
behavior as $J^\mu(\lambda)$ as $\lambda \rightarrow 0$, and thus it
has an analogous multipole series. In particular, as a distribution,
$T^M_{\mu \nu} (\lambda) \rightarrow 0$ as $\lambda \rightarrow 0$, whereas
to first order in $\lambda$, the matter stress-energy tensor takes the form
\begin{equation}
T^{M,(1)}_{\mu \nu}  \equiv \lim_{\lambda \rightarrow 0} \frac{\partial}
{\partial \lambda} T^{M}_{\mu \nu}(\lambda)
= \mathcal{T}^{M}_{\mu \nu}(t) \frac{\delta(x^i-z^i(t))}{\sqrt{-g}}\frac{d\tau}{dt} \,\, .
\label{eq:TM1}
\end{equation}

The electromagnetic stress energy tensor is given by eq.(\ref{eq:Tem}),
with $F_{\mu \nu} = F^{\textrm{ext}}_{\mu \nu} +
F^{\textrm{self}}_{\mu \nu}$. Thus, the total electromagnetic stress-energy
naturally breaks up into a sum of three terms,
\begin{equation}
T^{EM}_{\mu \nu} = T^{\textrm{ext}}_{\mu \nu}
+ T^{\textrm{cross}}_{\mu \nu} + T^{\textrm{self}}_{\mu \nu} \,\, ,
\end{equation}
where the notation should be self-explanatory. The stress-energy
tensor $T^{\textrm{ext}}_{\mu \nu}$ of the external field
$F^{\textrm{ext}}_{\mu \nu}$ behaves smoothly in $(\lambda, x^\mu)$
and thus has a straightforward perturbative expansion in
$\lambda$. Since $F^{\textrm{ext}}_{\mu \nu}$ is a solution to the
homogeneous Maxwell's equations, we also have $\nabla^\mu
T^{\textrm{ext}}_{\mu \nu}(\lambda) = 0$ for all $\lambda$, so conservation
of $T^{\textrm{ext},(n)}_{\mu \nu}$ holds at each order in perturbation theory.

The behavior of the stress-energy tensor $T^{\textrm{self}}_{\mu \nu}$
of the self-field $F^{\textrm{self}}_{\mu \nu}$ as $\lambda
\rightarrow 0$ is more delicate. Since $F^{\textrm{self},(0)}_{\mu \nu}
= 0$, one might expect that the lowest order contribution to
$T^{\textrm{self}}_{\mu \nu}$ would be obtained by substituting
$\lambda F^{\textrm{self},(1)}_{\mu \nu}$, eq.(\ref{eq:F1}), into
eq.(\ref{eq:Tem}). This would give an expression of the form $\lambda^2$
times the usual stress-energy tensor of a point charge. This would
suggest that the first order in $\lambda$ contribution to
$T^{\textrm{self}}_{\mu \nu}$ vanishes, and the second order
contribution is too singular on $\gamma$ to define a distribution. If
this were the case, then our analysis would be plagued by the same
type of infinite self-energy problems that occur in usual treatments
of point particles. Fortunately, as we now shall show, a more careful
treatment shows that to first order in $\lambda$,
$T^{\textrm{self}}_{\mu \nu}$ has a non-zero, well defined
distributional limit.

As previously mentioned in the Introduction, in Appendix
\ref{sec:app-scaling} we prove that $F^{\textrm{self}}_{\mu \nu}$ is
of the form (\ref{eq:Fform}), where $\tilde{F}_{\mu \nu}$ is a smooth
function of $(\lambda,t,[x^i-z^i(t)]/\lambda)$. We also prove in
Appendix \ref{sec:app-scaling} that if we define $\alpha = r =
\sqrt{\sum [x^i - z^i(t)]^2}$, define $\beta=\lambda/r$, and write
$\tilde{F}_{\mu \nu}$ as a function of $(\alpha, \beta, t, \theta,
\phi)$, then, at any fixed $(t, \theta, \phi)$, $\tilde{F}_{\mu \nu}$
is of the form of $\beta^2$ times a smooth function of $(\alpha,
\beta)$ at $(0,0)$. Consequently, $T^{\textrm{self}}_{\mu \nu}$ is of
the form
\begin{equation}
T_{\mu \nu}^{\textrm{self}}(\lambda,t,x^i) = \lambda^{-2}
\tilde{T}^{\textrm{self}}_{\mu \nu}(\lambda,t,[x^i-z^i(t)]/\lambda) \,\, ,
\label{Tselfform}
\end{equation}
where $\tilde{T}^{\textrm{self}}_{\mu \nu}$ is given in terms of $\tilde{F}_{\mu \nu}$
by eq.(\ref{eq:Tem}) and is a smooth function of
its arguments. Thus, $\tilde{T}^{\textrm{self}}_{\mu
  \nu}(\lambda)$ is of a form similar to that of
$J^\mu(\lambda)$ and $T^M_{\mu \nu} (\lambda)$ in our condition (i) of
the previous section, except that it is not of compact support in
$x^i$.  Nevertheless, from the properties of
$\tilde{F}_{\mu \nu}$, we know that at any fixed $(t, \theta,\phi)$,
$\tilde{T}^{\textrm{self}}_{\mu \nu}$ is of the form of $\beta^4$ times a smooth
function of $(\alpha, \beta)$.  If we view $T^{\textrm{self}}_{\mu \nu}(\lambda)$ for $\lambda > 0$ as a
distribution, the same manipulations as led to eq.(\ref{eq:manip}) now
yield
\begin{equation}
D_{T^{\textrm{self}}}[f] = \lambda \int \tilde{T}^{\textrm{self}}_{\mu\nu}(\lambda,t,\bar{x}^i) f^{\mu
  \nu} (t,z^i(t) + \lambda \bar{x}^i) \sqrt{-g}  dt d^3\bar{x} \,\, .
\label{Tdist}
\end{equation}
If $\tilde{T}^{\textrm{self}}_{\mu\nu}$ were of compact support in $\bar{x}^i$, we could
straightforwardly take the limit as $\lambda \rightarrow 0$ inside the
integral, as we did in eq.(\ref{eq:manip}). Nevertheless, using the
facts that (1) the integrand is smooth in $\bar{x}^i$, (2) the test
tensor field $f^{\mu \nu}$ is of compact support in $x^i$ (and, thus,
is of compact support in $\alpha$) and (3) for $\alpha$ in a compact
set, $\tilde{T}^{\textrm{self}}_{\mu \nu}$ is bounded by $C \beta^4$ for some constant
$C$, it is not difficult to show that for $\lambda \leq \lambda_0$
\begin{equation}
|\tilde{T}^{\textrm{self}}_{\mu\nu}(\lambda,t,\bar{x}^i)| |f^{\mu
  \nu} (t,z^i(t) + \lambda \bar{x}^i)|
\leq \frac{K}{\bar{r}^4 + 1}
\label{Tbnd}
\end{equation}
where $K$ is a constant (i.e., independent of $\lambda$ and
$x^i$). Since $1/[\bar{r}^4 + 1]$ is integrable with respect to
$\sqrt{-g} d^3\bar{x}$, the dominated convergence theorem then allows us to take
the limit as $\lambda \rightarrow 0$ inside the integral in this case
as well, and we obtain
\begin{equation}
D_{T^{\textrm{self}}}[f] = \lambda \int dt \frac{d\tau}{dt}
f^{\mu\nu}(t,z^i(t)) \int d^3\bar{x} \sqrt{g_3} \tilde{T}^{\textrm{self}}_{\mu\nu}(0,t,\bar{x}^i)
+ O(\lambda) \,\, ,
\label{Tdistlim}
\end{equation}
i.e., as distributions we have
\begin{align}
T_{\mu \nu}^{\textrm{self},(0)}& \equiv \lim_{\lambda \rightarrow 0}
T_{\mu \nu}^{\textrm{self}}(\lambda) = 0 \label{eq:T0} \\ T_{\mu \nu}^{\textrm{self},(1)}
& \equiv \lim_{\lambda \rightarrow 0} \frac{\partial} {\partial
  \lambda} T_{\mu \nu}^{\textrm{self}}(\lambda) = \mathcal{T}_{\mu \nu}^{\textrm{self}}(t) \frac{\delta(x^i-z^i(t))}{\sqrt{-g}}\frac{d\tau}{dt},\label{eq:T1}
\end{align}
with
\begin{equation}
\mathcal{T}_{\mu \nu}^{\textrm{self}}(t) \equiv \int
\tilde{T}_{\mu\nu}(0,t,\bar{x}^i) \sqrt{g_3} d^3\bar{x} \,\, .
\end{equation}
Thus, despite the fact that the electromagnetic self-field is not
confined to a world-tube at any $\lambda>0$, to first order in
$\lambda$, the electromagnetic self-stress-energy takes the form of a
$\delta$-function on $\gamma$. However, it should be noted that, in
contrast to the situation for $J^\mu$ and $T^M_{\mu \nu}$, at higher
orders in $\lambda$ the support of $T_{\mu \nu}^{\textrm{self}}$
cannot be confined to $\gamma$. Indeed, although we have shown above
that $D_{T^{\textrm{self}}}[f]$, eq.(\ref{Tdist}), is a $C^1$ function
of $\lambda$ at $\lambda = 0$, it does not appear that it is
smooth\footnote{A failure in obtaining a distributional series for
  $T_{\mu \nu}^{\textrm{self}}$ should not affect our analysis at
  higher orders, since $T_{\mu \nu}^{\textrm{self}}$ enters our
  equations only in the form $\nabla^\mu T_{\mu
    \nu}^{\textrm{self}}$. It is not difficult to show that
  $\nabla^\mu T_{\mu \nu}^{\textrm{self}}$ does have a distributional
  series with support on $\gamma$ of a form similar to that of $J^\mu$
  and $T^M_{\mu \nu}$.} in $\lambda$ at $\lambda = 0$.

There is nothing further that we can say beyond eqs.(\ref{eq:TM1}) and
(\ref{eq:T1}) about the individual forms of the material stress-energy
and electromagnetic self-stress-energy to first order in
$\lambda$.  However, defining the ``matter plus electromagnetic self-field'' stress-energy $T_{\mu \nu}$,
\begin{equation}\label{eq:T}
T_{\mu \nu} \equiv T^M_{\mu \nu} + T^{\textrm{self}}_{\mu \nu},
\end{equation}
we see from conservation of total stress-energy,
(\ref{eq:Tcons}), that for all $\lambda$ we have
\begin{eqnarray}
\nabla^\mu T_{\mu \nu}
&=& - \nabla^\mu T^{\textrm{ext}}_{\mu \nu} - \nabla^\mu
T^{\textrm{cross}}_{\mu \nu} \nonumber \\
&=& F^{\textrm{ext}}_{\mu \nu} J^\nu,
\end{eqnarray}
where we emphasize that $T_{\mu \nu}$, with no superscript, includes the matter and ``electromagnetic self-field'' contributions, as in equation \eqref{eq:T}.  The same type of calculation that led from eq.(\ref{eq:Jcons-dist}) to eq.(\ref{eq:J1-2}) then gives to first order in $\lambda$ that
\begin{equation}
T^{(1)}_{\mu \nu}(t) = m u_\mu u_\nu \frac{\delta(x^i-z^i(t))}{\sqrt{-g}}\frac{d\tau}{dt},
\label{eq:T1form}
\end{equation}
where $u^\mu$ is the unit tangent to $\gamma$, and $m$ is a constant (i.e.,independent of $t$).  In coordinates $(t,x^i)$ such that the constant-t hyperplanes are orthogonal to $u^\mu$, $m$ is given by
\begin{equation}
m = \int \left( \tilde{T}^{M}_{00} (\lambda=0,t,\bar{x}^i) + \tilde{T}^{\textrm{self}}_{00} (\lambda=0,t,\bar{x}^i) \right) \sqrt{g_3} \ d^3\bar{x}.
\label{eq:m}
\end{equation}
Furthermore, $u^\mu$ must satisfy
\begin{equation}
m u^\nu \nabla_\nu u_\mu
=  q u^\nu F^{\textrm{ext}}_{\mu \nu}(\lambda=0,t,z^i(t)) \,\, .
\label{eq:lorentzforce}
\end{equation}

We may summarize what we have just shown as follows: Consider any
one-parameter family of bodies satisfying the assumptions of the
previous section. Then, to first order in $\lambda$, the description
of the body is precisely that of a classical point charge/mass moving
on a Lorentz force trajectory of the external field. Specifically, the
charge-current of the body is given by eqs.(\ref{eq:J1-2}) and
(\ref{eq:q}), the sum of the body stress-energy and the
electromagnetic self-field stress-energy is given by
eq.(\ref{eq:T1form}), and the world line is given by
eq.(\ref{eq:lorentzforce}). It should be noted from eq.(\ref{eq:m})
that the electromagnetic self-energy contributes to the mass of the
body, i.e., there is a finite ``mass renormalization'' of the body
contributed by its electromagnetic field.  It should be emphasized
that the description of the body as a point charge/mass is an {\it
  output} of our calculations, that our derivation is mathematically
rigorous, and that no infinities have arisen in the derivation of our
formula (\ref{eq:m}) for the mass of the particle.

\section{Near-zone Limit and Perturbed Motion}
\label{sec:nearzone}

We now turn our attention to the first order in $\lambda$ corrections
to the motion. In order to do so, we must first address the issue of
what we mean by the ``corrections to motion''. As $\lambda \rightarrow
0$, the body shrinks down to a worldline $\gamma$, so at lowest order,
the motion is described by $\gamma$. As shown in the previous section,
$\gamma$ must satisfy the Lorentz force equation
(\ref{eq:lorentzforce}).  But, at any $\lambda > 0$, the body is of
finite size, so in order to find the ``correction'' to $\gamma$ at
finite $\lambda$, we would need to have a ``representative worldline''
to describe the motion of the body. It would be natural to take this
worldline to be the ``center of mass''. However, as we shall now
explain, it is far from obvious as to how to define a ``center of
mass''.

The difficulty arises from the fact that ``electromagnetic
self-energy'' must be included in a definition of center of
mass. Indeed, we have already seen in eq.(\ref{eq:m}) above that, at
lowest nonvanishing order, electromagnetic self-energy contributes to
the mass of the body. Since energy can be exchanged between the matter
and electromagnetic self-field---only their sum, $m$, is conserved at
leading order---it clearly would not make sense to omit the
electromagnetic self-energy from the definition of the center of
mass. However, if one defines the electromagnetic self-stress-energy $T_{\mu \nu}^{\textrm{self}}$ at finite $\lambda$ to be the electromagnetic stress-energy of the retarded solution $F_{\mu \nu}^{\textrm{self}}$, then the stress-energy of radiation that was emitted by the body in the distant past would be included.  Furthermore, integrals over the self-stress-energy defined this way---corresponding to that stress-energy's contributions to mass, center of mass, etc.---would diverge, on account of the slow, $1/r$ falloff of the retarded self-field.  It is therefore clear on physical and mathematical grounds that the full retarded self-field may not be used to provide a definition of center of mass.  On the other hand, it is far from obvious as to how one can exclude the energy of the ``emitted radiation'' while including the energy of the ``Coulomb self-field'', since one cannot make a clean distinction between the two at finite times.

Recently, one of us \cite{harte2009} has proposed an appropriate
definition of the ``center of mass'' for charged bodies that properly
takes into account the body's electromagnetic self-energy while not
including radiation emitted in the distant past.  It should be
possible to use the ``representative worldline'' provided by this
definition at small but finite $\lambda$ to define a notion of the
perturbative corrections to $\gamma$ to all orders in $\lambda$.
However, in this paper, we shall be concerned only with the leading
order correction to $\gamma$. As we shall see below, at this order,
there is a very simple and straightforward notion of ``center of
mass''---including self-field contributions---that can be used to
define the perturbed motion, and we will use this notion in our
analysis. It can be shown that this notion agrees with the much more
general definition of \cite{harte2009} to this order.\footnote{It was
  emphasized in \cite{harte2009} that there is a class of reasonable
  definitions of generalized momenta and hence center of mass that one
  may adopt.  The differences between the definitions begin at
  $O(\lambda^2)$ in the perturbation series discussed in this paper,
  which is precisely the order at which our type of definition would
  break down.}

Our analysis is most conveniently carried out in the ``near zone''
picture.  Therefore, in subsection \ref{subsec:nearzone} we introduce
the near-zone limit for the charge-current, stress-energy,
electromagnetic field, and metric, and we determine the properties of
the various near-zone fields.  In subsection \ref{subsec:parameters},
we present our definitions of the body parameters.  Finally, in
subsection \ref{subsec:motion} we re-derive the background Lorentz
force motion in the near-zone context, and then compute the perturbed
motion.

\subsection{Near-zone Limit and Properties}
\label{subsec:nearzone}

As analyzed in detail in the previous section, the ordinary
(``far-zone'') limit associated with our one-parameter family sees the
body shrink down in size to an effective point particle description.  Here we define a second, ``near-zone'' limit wherein the body remains at
fixed size.  To accomplish this, we measure spacetime distances (at
$\lambda > 0$) with a rescaled metric,
\begin{equation}\label{eq:gbar}
\bar{g}_{ab} \equiv \lambda^{-2} g_{ab} \,\,.
\end{equation}
The inverse metric $\bar{g}^{ab}$ is related by $\bar{g}^{ab} =
\lambda^2 g^{ab}$.  The Maxwell and matter equations, however, are not
satisfied relative to the barred metric $\bar{g}_{ab}$.  Therefore, we
also introduce rescaled fields,\footnote{Note that these
  rescalings are modified if indices are raised or lowered, since we
  use the barred metric to raise and lower indices for barred quantities and the unbarred metric to raise and lower indices
  for unbarred quantities.  For example, we have $\bar{J}_a = \lambda
  J_a$.}
\begin{align}
\bar{J}^{a} & \equiv \lambda^3 J^{a} \label{eq:Jbar} \\
\bar{T}^M_{ab} & \equiv T^M_{ab} \label{eq:Tmbar} \\
\bar{F}_{ab} & \equiv \lambda^{-1} F_{ab} \label{eq:Fbar},
\end{align}
so that the Maxwell and matter equations,
\begin{align}
\bar{\nabla}^b \bar{F}_{ab} & = 4 \pi \bar{J}_a \label{eq:max1-bar} \\
\bar{\nabla}_{[a}\bar{F}_{b]c} & = 0 \label{eq:max2-bar} \\
\bar{\nabla}_a \bar{J}^a & = 0 \label{eq:Jcons-bar} \\
\bar{\nabla}^b \left( \bar{T}^{B}_{ab} + \bar{T}^{EM}_{ab} \right) & = 0,\label{eq:Tcons-bar}
\end{align}
are satisfied.  Here the barred metric $\bar{g}_{ab}$ is used to raise
and lower indices, and $\bar{\nabla}_a$ is the derivative operator
associated with the barred metric (which, however, agrees with the
derivative operator associated with the unbarred metric).  The
electromagnetic stress-energy $\bar{T}^{EM}_{ab}$ is constructed from
$\bar{F}_{ab}$ in the usual way.

To complete the description of the near-zone limit, we introduce at
each time $t_0$ the ``scaled coordinates'' $(\bar{t},\bar{x}^i)$
defined by
\begin{equation}\label{eq:xbar}
\bar{t}\equiv\frac{t-t_0}{\lambda} \,,\,\,\,\,\,\,
\bar{x}^i\equiv\frac{x^i-z^i(t_0)}{\lambda}.
\end{equation}
The limit as $\lambda \rightarrow 0$ at fixed $(\bar{t},\bar{x}^i)$ of
a barred quantity in barred coordinates constitutes its
\textit{near-zone limit}.  Physically, as $\lambda \rightarrow 0$, an observer
at fixed $(\bar{t},\bar{x}^i)$ approaches the spacetime point
$(t=t_0,x^i=z^i(t_0))$ along with the shrinking body, but since the observer uses the rescaled metric to measure distances, the space and time intervals between the observer and the point $(t_0,z^i(t_0))$  remain finite. Thus the near-zone limit has the
interpretation of ``zooming in'' on the body (in space and time) at the same rate at which the body is shrinking.

In order investigate the properties of the near-zone limit, it is
convenient to specialize to coordinates where $z^i(t)=0$, and we shall
do so for the remainder of this paper.  In
this case, near-zone coordinate components are related to far-zone
coordinate components by simple expressions,
\begin{align}
\bar{g}_{\bar{\mu} \bar{\nu}}(\lambda,t_0;\bar{t},\bar{x}^i) & = g_{\mu \nu}(\lambda,t=t_0 + \lambda \bar{t},x^i=\lambda \bar{x}^i) \label{eq:gbar-comp}\\
\bar{J}^{\bar{\mu}}(\lambda,t_0;\bar{t},\bar{x}^i) & = \lambda^2
J^{\mu}(\lambda,t=t_0 + \lambda \bar{t},x^i=\lambda
\bar{x}^i) \label{eq:Jbar-comp} \\ \bar{T}^M_{\bar{\mu}
  \bar{\nu}}(\lambda,t_0;\bar{t},\bar{x}^i) & = \lambda^2 T^M_{\mu
  \nu}(\lambda,t=t_0 + \lambda \bar{t},x^i=\lambda
\bar{x}^i) \label{eq:Tbar-comp} \\ \bar{F}_{\bar{\mu}
  \bar{\nu}}(\lambda,t_0;\bar{t},\bar{x}^i) & = \lambda F_{\mu
  \nu}(\lambda,t=t_0 + \lambda \bar{t},x^i=\lambda
\bar{x}^i) \,\, .
\label{eq:Fbar-comp}
\end{align}
Note the presence of bars on the indices for the tensors on the left
side of these equations (denoting components in barred coordinates)
and the absence of bars on the indices of tensors on the right sides
(denoting corresponding components in the corresponding unbarred
coordinates). It then follows directly from our assumptions and
eq.~\eqref{eq:Fform} that
\begin{align}
\bar{J}^{\bar{\mu}}(\lambda,t_0;\bar{t},\bar{x}^i) & = \tilde{J}^{\mu}(\lambda,t_0 + \lambda \bar{t},\bar{x}^i) \label{eq:Jbar-tilde} \\
\bar{T}^M_{\bar{\mu}
  \bar{\nu}}(\lambda,t_0;\bar{t},\bar{x}^i) & = \tilde{T}^M_{\mu
  \nu}(\lambda,t_0 + \lambda \bar{t},\bar{x}^i) \label{eq:Tbar-tilde} \\
\bar{F}_{\bar{\mu} \bar{\nu}}(\lambda,t_0;\bar{t},\bar{x}^i) & = \tilde{F}_{\mu \nu}(\lambda,t_0 + \lambda \bar{t},\bar{x}^i) + \lambda F^{\textrm{ext}}_{\mu \nu}(\lambda, t_0 + \lambda \bar{t}, \lambda \bar{x}^i) \, \, . \label{eq:Fbar-tilde}
\end{align}
This gives the behavior of the near-zone fields in terms of the
(smooth) ``tilded'' quantities of our assumptions.
Equations (\ref{eq:Jbar-tilde})-(\ref{eq:Fbar-tilde}) make clear that in
the near-zone limit the size of the body remains finite (as measured
by the barred metric) and all properties of the body (as described by
the barred fields) go to well-defined, finite limits.  We also see
that the effects of
the external universe, as well as of the body's own future and past,
reduce in importance and finally disappear at $\lambda=0$.  This
contrasts with the far-zone limit analyzed in the previous section,
where the effects of the external universe go to well defined, finite
limits, but the effects of the body reduce in importance and finally
disappear at $\lambda = 0$.

Since the near-zone fields have now been seen explicitly to be smooth in
$\lambda,\bar{t},\bar{x}^i$, we know in particular that the near-zone
perturbation series exists to all orders in $\lambda$.  We will denote
the terms in this series in analogy with the notation used in the
far-zone limit: For a
generic barred quantity $\bar{f}$, we write
$\bar{f}^{(n)}(\bar{x}^\mu)=\frac{1}{n!}\lim_{\lambda \rightarrow
  0}\frac{\partial^n}{\partial \lambda^n}\bar{f}(\lambda,\bar{x}^\mu)$.
Thus it is
our convention that the superscript $(n)$ refers to $\lambda$-derivatives
at $\lambda = 0$
at fixed $\bar{x}^{\mu}$ when the superscripted quantity is barred; otherwise,
it refers to $\lambda$-derivatives at fixed $x^\mu$. Note that since the
charge-current and material stress-energy are of compact support in
$\bar{x}^i$ at all $\lambda$, this property also holds at each order in
perturbation theory.

It follows immediately from eqs.(\ref{eq:gbar-comp})-(\ref{eq:Fbar-comp}) that
each near-zone quantity $\bar{f}$ depends on $t_0$ and $\bar{t}$
only in the combination $t_0 + \lambda \bar{t}$, and that $\bar{f}$
must be a smooth function of $t_0 + \lambda \bar{t}$.  This has two
important consequences.  First, the dependence of any $n$th order
perturbative quantity $\bar{f}^{(n)}$ on $\bar{t}$ can be at most
$n$th order polynomial in $\bar{t}$
\begin{equation}\label{eq:time-dep}
\bar{f}^{(n)}(\bar{t},\bar{x}^i) = \sum_{m=0}^{n} A^{(n)}_m(\bar{x}^i) \bar{t}^m \,\,.
\end{equation}
In particular, all zeroth order near-zone quantities are stationary.
Second, we have a simple relationship between $\bar{t}$
and $t_0$ derivatives of $\bar{f}^{(n)}$,
\begin{equation}\label{eq:consistency}
\frac{\partial}{\partial \bar{t}}\bar{f}^{(n+1)} =
\frac{\partial}{\partial t_0}\bar{f}^{(n)}.
\end{equation}
Equations \eqref{eq:time-dep} and \eqref{eq:consistency} are satisfied
component-by-component by \textit{all} the near-zone quantities
$\bar{J}^{\bar{\mu}},\bar{T}^M_{\bar{\mu}\bar{\nu}},\bar{F}_{\bar{\mu}\bar{\nu}},\bar{g}_{\bar{\mu}\bar{\nu}}$.
Equation \eqref{eq:consistency} reflects the fact that the near-zone
perturbation series at each time $t_0$ are all defined with reference
to the single (``far-zone'') one-parameter-family.  Thus, perturbed
near-zone quantities at different scaling times $t_0$ cannot be
specified independently but instead must be related by equation
\eqref{eq:consistency}.  We refer to this relation here, as in we did for
the analogous condition
in \cite{gralla-wald}, as a ``consistency condition''.  We
will make use of the consistency condition in our calculations, below.

We now turn to the relationship between near-zone and far-zone
perturbative quantities. In the previous section, we saw that the
far-zone expansion of $J^\mu$ was distributional in nature, with the
$(n+1)$st order term, $J^{(n+1)\mu}$, in the expansion being described
as a sum of multipoles (i.e., derivatives of $\delta$-functions) up to
order $n$. It can be seen that at $t=t_0$, for $m \leq n$, the $m$th
order multipole (in the sense of an $m$th-derivative of a delta
function) appearing in $J^{(n+1)\mu}$ contributes to the $m$th
multipole moment (in the sense of a moment integral) of
$\bar{J}^{(n-m)\bar{\mu}}$ at $\bar{t}=0$.\footnote{Note, however,
  that the far-zone multipole series was, in effect, defined with
  $\gamma$ taken to be at the origin of coordinates, whereas near-zone
  moments will be defined below relative to a $\lambda$-dependent
  worldline, so the relationship between higher order far-zone and
  near-zone multipole moments will be complicated.}  Conversely,
 the multipole moments of the near-zone $p$th order charge-current $\bar{J}^{(p)\bar{\mu}}$ correspond to moments appearing at $p$th and higher order in the far-zone expansion of $J^\mu$.  This is characteristic of the kind of
``mixing of orders'' of quantities appearing in far-zone and near-zone
expansions. Similar
results hold for the near-zone and far-zone expansions of $T^M_{\mu
  \nu}$.

Consider, now, the near-zone external electromagnetic field
\begin{equation}
\bar{F}^{\textrm{ext}}_{\bar{\mu} \bar{\nu}}(\lambda,t_0;\bar{t},\bar{x}^i) \equiv \lambda F^{\textrm{ext}}_{\mu \nu}(\lambda, t_0 + \lambda \bar{t}, \lambda \bar{x}^i). \label{eq:Fbar-tilde-ext}
\end{equation}
Since $F^{\textrm{ext}}_{\mu \nu}$ is a smooth function of its
arguments, we see that the perturbative expansion of
$\bar{F}^{\textrm{ext}}_{\bar{\mu} \bar{\nu}}$ at order $n$ can
  depend only polynomially on $\bar{x}^i$, with the degree of the
  polynomial being no higher than $n-1$.  Explicitly, the first few terms
  are given by
\begin{align}
\bar{F}^{\textrm{ext},(0)}_{\bar{\mu} \bar{\nu}}(\lambda,t_0;\bar{t},\bar{x}^i) & = 0 \label{eq:Fbar-ext-0}\\
\bar{F}^{\textrm{ext},(1)}_{\bar{\mu} \bar{\nu}}(\lambda,t_0;\bar{t},\bar{x}^i) & = F^{\textrm{ext},(0)}_{\mu \nu}|_{t=t_0,x^i=0} \label{eq:Fbar-ext-1}\\
\bar{F}^{\textrm{ext},(2)}_{\bar{\mu} \bar{\nu}}(\lambda,t_0;\bar{t},\bar{x}^i) & = F^{\textrm{ext},(1)}_{\mu \nu}|_{t=t_0,x^i=0} + \bar{x}^i \partial_i F^{\textrm{ext},(0)}_{\mu \nu}|_{t=t_0,x^i=0} + \bar{t} \partial_0 F^{\textrm{ext},(0)}_{\mu \nu}|_{t=t_0,x^i=0} \label{eq:Fbar-ext-2}
\end{align}
These equations again show explicitly the characteristic ``mixing of
orders'' in the relationship between near-zone and far-zone fields.
For example, equation \eqref{eq:Fbar-ext-1} demonstrates that the
\textit{first}-order near-zone electromagnetic field is a constant (in
$\bar{t},\bar{x}^i$) whose value is given by the value of the
\textit{zeroth}-order far-zone field on the worldline at time $t_0$.
Notice also that the zeroth-order near-zone external electromagnetic
field vanishes, corresponding to the vanishing effects of the
``external universe'' in the near-zone limit.

We turn now to the near-zone expansion of the electromagnetic
self-field $F_{\mu \nu}^{\mathrm{self}}$. The far-zone expansion of
$F_{\mu \nu}^{\mathrm{self}}$ was given above in
eq.~(\ref{eq:F-series}). Writing $r = \lambda \bar{r}$ and $t = t_0 + \lambda \bar{r}$ in the first
line of eq.~(\ref{eq:F-series}), we obtain the corresponding near-zone
series (dropping the remainder terms for convenience)
\begin{align}
\bar{F}_{\bar{\mu} \bar{\nu}}^{\mathrm{self}}(\bar{t},\bar{r},\theta,\phi) & = \sum_{n=0}^N \sum_{m=0}^M \lambda^{n} \bar{r}^{n-m-2}(\mathcal{F}_{\mu \nu})_{nm}(t_0+\lambda \bar{t},\theta,\phi) \\
& = \sum_{n=0}^N \sum_{m=0}^M \sum_{p=0}^P \lambda^{n+p} \bar{t}^p \bar{r}^{n-m-2}(\mathcal{F}_{\mu \nu})_{nmp}(t_0)
\end{align}
where we have defined $(\mathcal{F}_{\mu \nu})_{nmp} \equiv
\frac{1}{p!}\left(\frac{\partial}{\partial t}\right)^p(\mathcal{F}_{\mu
  \nu})_{nm}|_{t=t_0}$ in the second line. Note that this expansion
gives useful information at {\it large} $\bar{r}$, in contrast to the
far-zone expansion eq.~(\ref{eq:F-series}), which gives useful
information at {\it small} $r$.  (Of course, these series contain precisely the same information, which is useful only at both small $\alpha=r=\lambda \bar{r}$ and small $\beta=\lambda/r=1/\bar{r}$.)  Note that in the near-zone series, at
$n$th order in $\lambda$ the highest possible combined positive powers
of $\bar{t}$ and $\bar{r}$ is $n$, but arbitrarily high inverse powers
of $\bar{r}$ can occur. Comparing with eq.~\eqref{eq:F1}, we can
explicitly evaluate the first few orders of the near-zone series for
$F_{\bar{\mu} \bar{\nu}}^{\mathrm{self}}$ at $\bar{t}=0$ as follows
\begin{align}
\bar{F}^{\textrm{self},(0)}_{i0}|_{\bar{t}=0} & = \frac{q}{\bar{r}^2}n_i + O\left(\frac{1}{\bar{r}^3}\right) \label{eq:Fbar-self-0-0i}\\
\bar{F}^{\textrm{self},(0)}_{ij}|_{\bar{t}=0} & = O\left(\frac{1}{\bar{r}^3}\right) \label{eq:Fbar-self-0-ij} \\
\bar{F}^{\textrm{self},(1)}_{i0}|_{\bar{t}=0} & = \frac{q}{\bar{r}} \left( - \frac{1}{2}(a_j n^j)n_i - \frac{1}{2} a_i \right) + O\left(\frac{1}{\bar{r}^2}\right) \label{eq:Fbar-self-1-0i} \\
\bar{F}^{\textrm{self},(1)}_{ij}|_{\bar{t}=0} & = O\left(\frac{1}{\bar{r}^2}\right) \label{eq:Fbar-self-1-ij} \\
\bar{F}^{\textrm{self},(2)}_{i0}|_{\bar{t}=0} & = q \left( \frac{3}{8} (a_j n^j)^2 n_i + \frac{3}{4} (a_j n^j) a_i + \frac{1}{8} a_j a^j n_i + \frac{1}{2} \dot{a}_0 n_i + \frac{2}{3} \dot{a}_i \right) +O\left(\frac{1}{\bar{r}}\right) \label{eq:Fbar-self-2-0i} \\
\bar{F}^{\textrm{self},(2)}_{ij}|_{\bar{t}=0} & = -q\frac{1}{2}\dot{a}_{[i}n_{j]} + O\left(\frac{1}{\bar{r}}\right) \,\, , \label{eq:Fbar-self-2-ij}
\end{align}
where, in this equation and all following equations in this paper, we
drop the bars on the coordinate components of barred quantities, i.e.,
it is understood that the components of any barred quantity is taken
in barred coordinates.

\subsection{Center of Mass and Body Parameters}
\label{subsec:parameters}
In order to determine the corrections to Lorentz force motion, we
would like to define a notion of the ``center of mass worldline'' of
the body at finite $\lambda$. However, as discussed at the beginning of
this section, this is a highly nontrivial notion for a charged body,
since it is far from obvious how self-field contributions to the
center of mass should be included. Fortunately, in order to define the first order
correction to Lorentz force motion in the far zone, we only need to define
a notion of center of mass to zeroth order in the near zone. As we shall now see,
this can be done in a very simple and straightforward manner.
However,
if we wished to calculate higher order corrections, we would need to
employ a much more sophisticated notion of ``center of mass'', such as
given in \cite{harte2009}.

To define the center of mass to first order in $\lambda$, consider an
arbitrary smooth, one-parameter family of worldlines $\gamma(\lambda)$
such that $\gamma(0)$ is the worldline $\gamma$ of condition (i) of
section II. Choose Fermi normal coordinates about $\gamma(\lambda)$,
corresponding to working in the ``rest frame'' of the body.  In these
coordinates, the (flat) spacetime metric takes the form
(see,e.g.,\cite{poisson})
\begin{equation}\label{eq:g-comp}
\begin{split}
g_{00} & = - 1 - 2 a_i(\lambda,t) x^i - (a_i(\lambda,t) x^i)^2 + O(r^3) \\
g_{i0} & = O(r^3) \\
g_{ij} & = \delta_{ij} \,\, ,
\end{split}
\end{equation}
where $a^i(\lambda,t)$ is the acceleration of $\gamma(\lambda)$ at time $t$.
The components of the corresponding scaled metric are given by
\begin{equation}\label{eq:gbar-series}
\begin{split}
\bar{g}_{00} & = - 1 - 2 \lambda a_i(t_0) \bar{x}^i - \lambda^2 \left[ 2\dot{a}_i(t_0) \bar{x}^i \bar{t} + (a_i(t_0) \bar{x}^i)^2 + 2 \delta a_i(t_0) x^i \right] + O(\lambda^3) \\
\bar{g}_{i0} & = O(\lambda^3) \\
\bar{g}_{ij} & = \delta_{ij} \,\, ,
\end{split}
\end{equation}
where we have defined $a^i(t) \equiv a^i(\lambda=0,t)$ and $\delta a^i
\equiv \partial_\lambda a^i(\lambda=0,t)$, and the overdot denotes the
derivative with respect to $t$.  Thus, $a^i$ corresponds to the
(far-zone) acceleration of the unperturbed worldline $\gamma(0)$, and
$\delta a^i$ corresponds to the perturbed (far-zone) acceleration of
$\gamma(\lambda)$ to first order in $\lambda$. In accordance with the
conventions stated at the end of the previous subsection, we have
dropped the bars on the coordinate components of
$\bar{g}_{\bar{\mu}\bar{\nu}}$ on the left side of these equations.

As before, we define the total body stress-energy tensor $\bar{T}_{ab}$ by
\begin{equation}
\bar{T}_{ab} \equiv \bar{T}^{M}_{ab} + \bar{T}^{\textrm{self}}_{ab} \,\, .
\end{equation}
At zeroth order in the near-zone, the spacetime metric
(\ref{eq:gbar-series}) is Minkowskian and the total body stress-energy
is stationary (since all zeroth order near-zone quantities are
stationary). Therefore, it is natural to define
the total body mass at zeroth order in the near-zone by
\begin{equation}\label{eq:mass}
m(t_0) \equiv \int \bar{T}^{(0)}_{00} d^3 \bar{x}.
\end{equation}
Since $\bar{F}^{\textrm{self},(0)}_{\mu \nu}$ is a
stationary, regular solution of Maxwell's equations that falls off at
large $\bar{r}$ as $1/\bar{r}^2$, the zeroth order self-field
stress-energy falls off as $1/\bar{r}^4$, so the integral defining $m$
converges.  Indeed, it is easily seen that $m$ agrees with the total
body mass at first order in the far-zone, eq.~\eqref{eq:m}, which we
also denoted as $m$.  Therefore, in particular, $m$ is independent of
$t_0$, a fact we will confirm in the next subsection by our
near-zone analysis.

We define the center of mass, $\bar{X}^i_{\rm CM} $, to zeroth order
in the near zone by,
\begin{equation}
\bar{X}^i_{\rm CM}(t_0) = \frac{1}{m} \int \bar{T}^{(0)}_{00}\bar{x}^i d^3\bar{x}
\end{equation}
The integrand in this equation falls off only as $1/\bar{r}^3$, so the integral defining $\bar{X}^i_{\rm CM} $ does not converge absolutely. However, the $1/\bar{r}^4$ part of $ \bar{T}^{(0)}_{00}$ (computed from equation \eqref{eq:Fbar-self-0-0i}) is spherically symmetric, so the angular average of the $1/\bar{r}^3$ part of the integrand vanishes. Thus, the integral defining $\bar{X}^i_{\rm CM} $ is well defined as a limit as $\bar{R} \rightarrow \infty$ of the integral
over a ball of radius $\bar{R}$.

The center of mass changes under a change of the origin of the near-zone
coordinates
$\bar{x}^i$.  This corresponds to a first-order in $\lambda$ change of
far-zone origin $x^i$, or, equivalently, to a change in the choice of
$\gamma(\lambda)$ to first-order in $\lambda$ in the far-zone.  We
therefore may define the \textit{perturbed} motion
to first-order in $\lambda$ in the far-zone by the demand that
the \textit{zeroth} order near-zone center of mass vanish,
\begin{equation}
\bar{X}^i_{\rm CM}(t_0) = 0 \,\, .
\label{eq:CM}
\end{equation}
Since $\bar{X}^i_{\rm CM}$ is defined relative to the Fermi (i.e.,
``rest frame'') coordinate components of the stress-energy, this
demand corresponds to the requirement that an observer moving along
the worldline assign his own position as the center of mass of the
body at each time $t_0$. In the next subsection, we shall see by
direct computation that this condition gives rise to an equation satisfied by
the perturbed acceleration $\delta a^i = \partial_\lambda a^i
|_{\lambda=0}$, thereby defining a correction to Lorentz force motion.

We define the \textit{spin tensor} $S^{\mu \nu}$ to zeroth order in
the near zone by
\begin{equation}
S^{0j}(t_0) = - S^{j0}(t_0) = \int \bar{T}^{(0)00}\bar{x}^j d^3\bar{x}
\end{equation}
and
\begin{equation}\label{eq:spin}
S^{ij}(t_0) \equiv 2 \int \bar{T}^{(0)i}_{\ \ \ \ 0}\bar{x}^j d^3\bar{x}.
\end{equation}
Thus, when the center of mass
condition is imposed, we have
\begin{equation}
S^{0j} = 0 \,\, .
\label{spinCM}
\end{equation}
We shall assume that the center of mass condition has been imposed
in the following. As we
shall see from eq.(\ref{eq:Ti0k}) below, it follows from lowest order
conservation of stress-energy that $S^{ij}$ is antisymmetric, $S^{ij}
= - S^{ji}$.  Note that the integral defining $S^{ij}$ converges
absolutely, since $\bar{F}^{\textrm{self},(0)}_{0i}$ falls off as
$1/\bar{r}^2$ and $\bar{F}^{\textrm{self},(0)}_{ij}$ falls off as
$1/\bar{r}^3$.  The spin vector $S_i$ is defined by
\begin{equation}
S_i = \frac{1}{2} \epsilon_{ijk} S^{jk} \,\, .
\end{equation}

To first order in the near-zone, the spacetime metric
(\ref{eq:gbar-series}) is no longer Minkowskian, but it is stationary,
with timelike Killing field $\partial/\partial \bar{t}$. The
stress-energy current $\bar{T}_{\mu \nu} (\partial/\partial
\bar{t})^\nu$ is therefore conserved to first order,
and it is natural to define the
first order correction to the mass, $\delta m$, by
\begin{equation}
\delta m(t_0) \equiv \int_{\Sigma} \frac{\partial}{\partial \lambda} \left. \left( \bar{T}_{ab}
(\frac{\partial}{\partial \bar{t}})^b d \Sigma^a \right) \right|_{\lambda=0} \,\, ,
\end{equation}
where the integral is independent of the spacelike hypersurface $\Sigma$.
This yields
\begin{equation}
\delta m(t_0) = \int \left. \bar{T}^{(1)}_{00}\right|_{\bar{t}=0}
d^3 \bar{x} \,\, ,
\label{eq:delta-M}
\end{equation}
where the center of mass condition has been used.  Since the
stress-energy $\bar{T}^{(1)}_{00}$ decays only as $1/\bar{r}^3$ on
account of the ``cross term'' contribution of
$\bar{F}^{\textrm{self},(0)}_{\bar{\mu} \bar{\nu}}$ and
$\bar{F}^{\textrm{self},(1)}_{\bar{\mu} \bar{\nu}}$, this integral
does not converge absolutely.  However, the explicit asymptotic form
of $\bar{T}^{(1)}_{00}$ (computed from equations
(\ref{eq:Fbar-self-0-0i}-\ref{eq:Fbar-self-1-ij})) is
\begin{equation}
\left. \bar{T}^{(1)}_{00}\right|_{\bar{t}=0} = -\frac{1}{4 \pi}\frac{q^2}{\bar{r}^3} a_j n^j + O\left(\frac{1}{\bar{r}^4}\right) \,\, , \label{eq:Tem1-00}
\end{equation}
from which it can be seen that the angular average of the
$1/\bar{r}^3$ part vanishes, so that the integral defining $\delta m$
is well defined as a limit as $\bar{R} \rightarrow \infty$ of the
integral over a ball of radius $\bar{R}$. In view of eq.~(\ref{eq:delta-M}), it would
be natural to interpret $\delta m$ as the first order correction to the rest mass
of the body (including its electromagnetic self-energy).

The total charge $q(\lambda)$ is given at finite $\lambda$ by
\begin{equation}\label{eq:q-exact}
q(\lambda) = \int_{\Sigma} J^a d \Sigma_a,
\end{equation}
and is independent of the choice of hypersurface $\Sigma$.
To zeroth order in the near-zone, the charge $q$ is therefore given by
\begin{equation}
q  \equiv \int \bar{J}^{(0)0} d^3 \bar{x}
\end{equation}
It is not difficult to see that this agrees with the first order
charge in the far-zone, which we previously also denoted as $q$.

The zeroth order electromagnetic dipole moment tensor $Q^{\mu \nu}$
of the body is defined by
\begin{equation}
\label{eq:Q}
Q^{\mu j}(t_0) \equiv \int \bar{J}^{(0) \mu} \bar{x}^j d^3\bar{x}\,\, ,
\end{equation}
together with $Q^{j 0} = - Q^{0 j}$. We shall see in the next
subsection that the purely spatial components $Q^{ij}$ are
automatically antisymmetric by lowest-order conservation of charge, so
we have $Q^{\mu \nu} = Q^{[\mu \nu]}$. The time-space components of $Q^{\mu
  \nu}$ correspond to the electric dipole moment
\begin{equation}
p^i = Q^{0i}
\end{equation}
whereas the purely spatial components of $Q^{\mu \nu}$ correspond to the magnetic
dipole moment
\begin{equation}
\mu_i = -\frac{1}{2}\epsilon_{ijk} Q^{jk} \,\, .
\end{equation}

By eq.\eqref{eq:q-exact}, we see that the first
order near-zone correction, $\delta q$,
to the charge is given by
\begin{equation}
\label{eq:delta-q}
\delta q  \equiv \int \bar{J}^{(1)0}|_{\bar{t}=0} d^3\bar{x} + a^i Q_{0i} \,\, .
\end{equation}
Note that the analog of the last term in this equation was absent from
eq.~(\ref{eq:delta-M}) on account of the imposition of the center of
mass condition. Since $q(\lambda)$ is independent of $t_0$ for all $\lambda$, it
follows immediately that $\delta q$ is independent of $t_0$.

\subsection{Derivation of Motion}
\label{subsec:motion}

We now turn to the
derivation of the perturbed equations of motion, which will be obtained
directly from
conservation of stress-energy, eq.~\eqref{eq:Tcons-bar}.
Writing $\bar{T}_{ab} = \bar{T}^{M}_{ab}+\bar{T}^{\textrm{self}}_{ab}$
as before, we obtain from
eq.~\eqref{eq:Tcons-bar}
\begin{equation}\label{eq:conv-cons-barred}
\bar{\nabla}^{b} \bar{T}_{ab} = \bar{J}^{b} \bar{F}^{\textrm{ext}}_{ab}.
\end{equation}

At zeroth order in the near-zone, the near-zone metric is flat
(see eq.~\eqref{eq:gbar-series}), the external electromagnetic field
vanishes (see eq.~(\ref{eq:Fbar-ext-0})), and
all zeroth order near-zone quantities are stationary (see
eq.~\eqref{eq:time-dep}). Therefore,
at zeroth order, we obtain
\begin{equation}
\label{eq:Tcons-bar0}
\bar{\partial}^j \bar{T}^{(0)}_{\mu j} = 0,
\end{equation}
where we again remind the reader that we have dropped bars on indices
of barred quantities. By multiplying the time and space components of
this equation by $x^i$ and integrating over space, we obtain
\begin{align}
\label{eq:Ti0}
\int \bar{T}^{(0)}_{i0} d^3 \bar{x} & = 0 \\
\label{eq:Tij}
\int \bar{T}^{(0)}_{ij} d^3 \bar{x} & = 0
\end{align}
Similarly, by multiplying the time and space components
by $x^i x^k$ and integrating over space, we obtain
\begin{align}
\label{eq:Ti0k}
\int \bar{T}^{(0)i}_{\ \ \ \ 0} \bar{x}^k d^3 \bar{x} & = - \int \bar{T}^{(0)k}_{\ \ \ \ 0} \bar{x}^i d^3 \bar{x} \\
\label{eq:Tijk}
\int \bar{T}^{(0)}_{ij} \bar{x}^k d^3 \bar{x} & = 0.
\end{align}
Equation~\eqref{eq:Ti0k} demonstrates the previously claimed
anti-symmetry of the (lowest-order) spin of the body, defined by
eq.~\eqref{eq:spin}. Equation~\eqref{eq:Tijk} follows from an
analogous anti-symmetry in $i$ and $k$, combined with the symmetry of
$T_{ij}$ in $i$ and $j$.

At zeroth order in the near-zone, conservation of charge-current yields
\begin{equation}
\bar{\partial}_i \bar{J}^{(0)i} = 0 \,\, ,
\end{equation}
from which we can analogously derive the useful relations,
\begin{align}
\int \bar{J}^{(0)i} d^3\bar{x} & = 0 \label{eq:Ji} \\ \int
\bar{J}^{(0)i} \bar{x}^j d^3 \bar{x} & = - \int \bar{J}^{(0)j}
\bar{x}^i \,\, .\label{eq:Jij}
\end{align}
The second of these equations demonstrates the previously claimed
anti-symmetry of the spatial components of the electromagnetic dipole
tensor $Q^{\mu \nu}$ defined by eq.\eqref{eq:Q}.

At first-order in the near-zone, we see the first appearances of the
acceleration (via the metric components \eqref{eq:gbar-series}) and
the external field (via equation \eqref{eq:Fbar-ext-1}). At first
order, conservation of stress-energy, eq.(\ref{eq:conv-cons-barred}),
explicitly yields
\begin{align}
\label{eq:Tcons-bar1-0}
\bar{\partial}^0\bar{T}^{(1)}_{00} + \bar{\partial}^i\bar{T}^{(1)}_{0i} + a^i \bar{T}^{(0)}_{0i} - \bar{J}^{(0)\mu}F^{\textrm{ext}}_{0\mu} & = 0 \\
\label{eq:Tcons-bar1-i}
\bar{\partial}^0\bar{T}^{(1)}_{k0} + \bar{\partial}^j\bar{T}^{(1)}_{kj} + a_k \bar{T}^{(0)}_{00} + a^j \bar{T}^{(0)}_{kj} - \bar{J}^{(0)\mu}F^{\textrm{ext}}_{k\mu} & = 0
\end{align}
where $a_i$ is evaluated at time $t=t_0$, and we have used
eq.~(\ref{eq:Fbar-ext-1}) to replace $\bar{F}^{\textrm{ext},(1)}_{\mu \nu}$
by $F^{\textrm{ext}}_{\mu \nu} \equiv F^{\textrm{ext},(0)}_{\mu
  \nu}|_{t=t_0,x^i=0}$.

We shall now use these equations to derive evolution equations for the
mass and spin to lowest order, as well as to determine the lowest order motion.
For the evolution of mass, we integrate
equation \eqref{eq:Tcons-bar1-0} over space (with volume element $d^3
\bar{x}$). Using the consistency condition \eqref{eq:consistency},
the first term evaluates to $-dm(t_0)/dt_0$.  The second term vanishes
by integration by parts, since $\bar{T}^{(1)}_{0i}$ falls off like
$1/\bar{r}^4$.  The third term vanishes by eq.~\eqref{eq:Ti0}, and the
fourth term vanishes by eq.(\ref{eq:Ji}).  Thus we obtain
\begin{equation}
\frac{d}{d t_0} m = 0 \,\, ,
\end{equation}
in agreement with what we already found in our far-zone analysis
given in section III.

For the evolution of spin, we multiply eq.~\eqref{eq:Tcons-bar1-i} by
$\bar{x}^i$ and integrate over space.  Again by the
consistency condition the first term evaluates to
$-(1/2)dS_{ik}(t_0)/dt_0$.  The third term vanishes by the center of mass
condition \eqref{eq:CM}, and the fourth term vanishes by eq.~\eqref{eq:Tijk}.
However, the second and fifth terms do not vanish, yielding
\begin{equation}\label{eq:flork}
\frac{1}{2}\frac{d}{d t_0} S_{ik} = \int \bar{T}^{(1)}_{ik}
d^3\bar{x} + Q^{\mu}_{\ i}F^{\textrm{ext}}_{k\mu}.
\end{equation}
The antisymmetric part of this equation provides the desired evolution
equation for the spin,
\begin{equation}
\label{eq:spin-evolution}
\frac{d}{d t_0} S_{ij} = 2 Q^{\mu}_{\ [i}F^{\textrm{ext}}_{j]\mu} \,\, .
\end{equation}
The right side of this equation corresponds to the usual formula for
the torque on an electromagnetic dipole. It also
should be noted that, since we work in Fermi
normal coordinates, $dS_{ik}/dt_0$ corresponds to the ``Fermi derivative'' of
$S_{ik}$, so eq.~(\ref{eq:spin-evolution}) automatically includes the
``Thomas precession'' (see eq.\eqref{eq:spin-cov} below).
The symmetric part of \eqref{eq:flork} yields the useful relation
\begin{equation}
\label{eq:T1ij} \int \bar{T}^{(1)}_{ij} d^3\bar{x} = Q^{\mu}_{\ (i}F^{\textrm{ext}}_{j)\mu}.
\end{equation}

To determine the lowest order motion, we
integrate equation \eqref{eq:Tcons-bar1-i} over space.
The first term vanishes by the consistency condition and eq.~\eqref{eq:Ti0},
the second term vanishes
since $\bar{T}^{(1)}_{ij}$ falls off as
$1/\bar{r}^3$, and the fourth term vanishes by eq.~\eqref{eq:Tijk}.
The third and fifth terms yield,
\begin{equation}
m a_i = q F^{\textrm{ext}}_{i0} \,\, .
\label{LF2}
\end{equation}
Thus, we reproduce the Lorentz force law previously derived by the far-zone
analysis of section III.

Finally, we derive an additional relation from first-order
stress-energy conservation by multiplying eq.~\eqref{eq:Tcons-bar1-0}
by $\bar{x}^i$ and integrating over space. The first term vanishes by
the consistency condition \eqref{eq:consistency} and the center of
mass condition \eqref{eq:CM}.  The second term may be integrated by
parts; the associated surface term vanishes because
$\bar{T}^{(1)}_{0i}$ falls off as $1/\bar{r}^4$.  The third and fourth
terms are simply expressed in terms of the spin \eqref{eq:spin} and
electromagnetic dipole \eqref{eq:Q} tensors. We obtain
\begin{equation}
\label{eq:T1i0}
\int \bar{T}^{(1)j}_{\ \ \ \ 0} d^3\bar{x} = \frac{1}{2}a_i S^{ij} + Q^{ij} F^{\textrm{ext}}_{i0} \,\, .
\end{equation}

In a similar manner, at first-order in the near-zone, conservation of
charge-current yields
\begin{equation}\label{eq:Jcons-bar1}
\bar{\partial}_0 \bar{J}^{(1)0} + \bar{\partial}_j \bar{J}^{(1)j} + a^{(0)}_j \bar{J}^{(1)j} = 0.
\end{equation}
Multiplying by $\bar{x}^i$ and integrating over space, we obtain
\begin{equation}
\label{eq:J1i}
\int \bar{J}^{(1)i} d^3 \bar{x} = \frac{d}{d t_0}Q^{0i} + a_j Q^{ji} \,\, .
\end{equation}

We now shall derive an equation for the perturbed acceleration $\delta a^i$
(see eq.~\eqref{eq:gbar-series}),
as well as an evolution equation for the perturbed mass, $\delta m$
(see eq.~(\ref{eq:delta-M})). At second-order, conservation of stress-energy
yields
\begin{align}
0 & = \bar{\partial}^0 \bar{T}_{00}^{(2)} + \bar{\partial}^i\bar{T}^{(2)}_{0i} + a^i T^{(1)}_{0i} + \delta a^i T^{(0)}_{0i} + \dot{a}^i \bar{T}_{0i}^{(0)}\bar{t} + 2 a_i \bar{x}^i \bar{\partial}_0 \bar{T}^{(1)}_{00} + 2 \dot{a}_i \bar{x}^i \bar{T}^{(0)}_{00} - a^i \bar{x}^i a^j \bar{T}^{(0)}_{0j} \nonumber \\ & - \bar{J}^{(0)\mu} \left(  \bar{x}^i \partial_i F^{\textrm{ext}}_{0\mu}+ \dot{F}^{\textrm{ext}}_{0\mu} \bar{t} + \delta F^{\textrm{ext}}_{0\mu} \right) - \bar{J}^{(1)\mu} F^{\textrm{ext}}_{0\mu} \label{eq:Tcons-bar2-0}\\
0 & = \bar{\partial}_0\bar{T}^{(2)}_{k0} + \bar{\partial}^i\bar{T}^{(2)}_{ki} + a_k \bar{T}^{(1)}_{00} + a^i\bar{T}^{(1)}_{ki} + \delta a_k \bar{T}^{(0)}_{00} + \delta a^i\bar{T}^{(0)}_{ki} + \dot{a}_k \bar{T}^{(1)}_{00} \bar{t} + \dot{a}^i\bar{T}^{(1)}_{ki} \bar{t} \nonumber \\ & + 2 a_i \bar{x}^i \bar{\partial}_0 \bar{T}^{(1)}_{k0} + \dot{a}_i \bar{x}^i T^{(0)}_{k0} - a_i \bar{x}^i a^j \bar{T}^{(0)}_{kj} - 3 a_k a_i \bar{x}^i \bar{T}^{(0)}_{00} \nonumber \\ & - \bar{J}^{(0)\mu} \left( \bar{x}^i \partial_i F^{\textrm{ext}}_{k\mu} + \dot{F}^{\textrm{ext}}_{k\mu} \bar{t} + \delta F^{\textrm{ext}}_{k\mu}\right) - \bar{J}^{(1)\mu} F^{\textrm{ext}}_{k\mu},
\label{eq:Tcons-bar2-i}
\end{align}
where we have used eqs.~\eqref{eq:gbar-series} and
(\ref{eq:Fbar-ext-2}) and have written $\delta F^{\textrm{ext}}_{\mu
  \nu} \equiv F^{\textrm{ext},(1)}_{\mu \nu}|_{t=t_0,x^i=0}$.  To
derive an equation for the perturbed acceleration $\delta a^i$, we
evaluate \eqref{eq:Tcons-bar2-i} at $\bar{t}=0$ and integrate over
space, applying the consistency condition \eqref{eq:consistency} to
various terms, using the center of mass condition \eqref{eq:CM}, using
eqs.~\eqref{eq:Tij} and \eqref{eq:Tijk}, and using the definitions of
the spin and dipole tensors given in section
\ref{subsec:parameters}. We obtain
\begin{align}
0 & = \frac{d}{d t_0} \int \bar{T}^{(1)}_{0k} d^3\bar{x} + \int
\bar{r}^2 n^i \bar{T}^{(2)}_{ki} d\Omega + a_k \delta m + a^i \int
T^{(1)}_{ki} d^3\bar{x} + \delta a_k m + a^i \frac{d}{dt_0}S_{ki} +
\frac{1}{2} \dot{a}^i S_{ki} \nonumber\\ & - Q^{\mu i} \partial_i
F^{\textrm{ext}}_{k \mu} - \delta F^{\textrm{ext}}_{k \mu} \int
\bar{J}^{(0)\mu} d^3\bar{x} - \int \bar{J}^{(1)\mu}F^{\textrm{ext}}_{k
  \mu} d^3 \bar{x} \,\,.
\label{2eom}
\end{align}
The second term in this equation is the surface term
arising from integration by parts of the second term in
eq.~(\ref{eq:Tcons-bar2-i}).  To evaluate this term, we compute the
asymptotic form of $\bar{T}^{(2)}_{ij}$ from equations
(\ref{eq:Fbar-self-0-0i}-\ref{eq:Fbar-self-2-ij}), obtaining
\begin{align}
\left. \bar{T}^{(2)}_{ij}\right|_{\bar{t}=0} & = -\frac{1}{4
  \pi}\frac{q^2}{\bar{r}^2} \bigg [ n_i n_j \left( 6 (a_k n^k)^2 +
  \frac{1}{4} a_k a^k + \dot{a}_0 \right) + \delta_{ij} \left( 4 (a_k
  n^k)^2 + \frac{1}{4} a_k a^k + \frac{1}{2} \dot{a}_0 + \frac{2}{3}
  \dot{a}_k n^k \right) \nonumber \\ & \qquad \ \quad \quad + 4 a_k
  n^k a_{(i}n_{j)} + \frac{4}{3} \dot{a}_{(i}n_{j)} + \frac{1}{4} a_i
  a_j \bigg ] + O\left(\frac{1}{\bar{r}^3}\right).
\end{align}
Since the
integral over a sphere of the product of an odd number of
normal vectors, $n^i$, vanishes, only the terms in
$\bar{T}^{(2)}_{ij}$ that are proportional to $\dot{a}^i$ survive when
computing the second term in eq.~(\ref{2eom}). We obtain
\begin{equation}
\int \bar{r}^2 n^i \bar{T}^{(2)}_{ki} d\Omega = \frac{2}{3} q^2 \dot{a}^i
\end{equation}
which can be recognized as the Abraham-Lorentz-Dirac self-force term (see
eq.~(\ref{eq:ALD-intro}) above).

Returning to eq.~(\ref{2eom}), we evaluate the first term using
eq.~\eqref{eq:T1i0}, we evaluate the fourth term using
eq.~\eqref{eq:T1ij}, and we use eqs.~\eqref{eq:Ji},
\eqref{eq:delta-q}, and \eqref{eq:J1i} to re-express the final two
terms in terms of the electromagnetic dipole and the corrected charge. After some algebra, we obtain
\begin{equation}
\label{eq:EOM}
m \delta a_i = -(\delta m) a_i + (\delta q) F^{\textrm{ext}}_{i0} + q
\delta F^{\textrm{ext}}_{i0} + \frac{2}{3} q^2 \dot{a}_i - \frac{1}{2}
Q^{\mu \nu}\partial_i F^{\textrm{ext}}_{\mu \nu} + \frac{d}{d
  t_0}\left(a^j S_{ji} - 2 Q^j_{\ [i}F^{\textrm{ext}}_{0]j}\right),
\end{equation}
where we have used Maxwell's equation
$\partial_{[\mu}F^{\textrm{ext}}_{\nu \rho]}=0$ to reexpress the fifth term on
the right side.

We also can
derive an evolution equation for the perturbed mass, $\delta m$, by
integrating equation \eqref{eq:Tcons-bar2-0} (evaluated at
$\bar{t}=0$) over $\bar{x}^i$, and performing similar manipulations as
above. The result is,
\begin{equation}\label{eq:mass-evolution}
\frac{d}{d t_0} \delta m = \frac{1}{2} Q^{\mu \nu}\partial_{0}F^{\textrm{ext}}_{\mu \nu} - \frac{d}{d t_0} \left( Q^{\mu 0} F^{\textrm{ext}}_{\mu 0} \right).
\end{equation}

Equation~(\ref{eq:EOM}) is our desired equation for the perturbed
acceleration $\delta a^i$. As described at the beginning of the next
section, this equation can be written as an evolution equation for the
deviation vector, $X^a$, describing the perturbation from Lorentz
force motion. Appearing on the right side of eq.~(\ref{eq:EOM}) are
the lowest order mass, $m$, the lowest order charge, $q$, the lowest order
acceleration $a^i$, the perturbed mass, $\delta m$, the perturbed
charge, $\delta q$, the spin tensor, $S^{\mu \nu}$, and the
electromagnetic dipole tensor, $Q^{\mu \nu}$. As previously derived,
$m$, $q$, and $\delta q$ are conserved, and $a^i$ is given by the
Lorentz force equation (\ref{LF2}). The evolution of $S^{\mu \nu}$ is
given by eq.~\eqref{eq:spin-evolution}, and the evolution of $\delta
m$ is given by eq.~(\ref{eq:mass-evolution}).  However, we cannot
obtain an evolution equation for the electromagnetic dipole tensor
$Q^{\mu \nu}$ because the behavior of this quantity is
``non-universal''. Indeed, it seems clear that by use of a ``Maxwell
demon''\footnote{Of course, the stress-energy of the Maxwell demon
  must be included in $T^M_{\mu \nu}$.}, we could make $Q^{\mu \nu}$
evolve with time any way that we wish.  Therefore, additional
conditions/assumptions beyond the Maxwell and matter equations
\eqref{eq:max1}-\eqref{eq:Jcons} and \eqref{eq:Tcons}
would need to be adjoined to the equations we have derived in order to
get deterministic evolution at this order.  In appendix
\ref{sec:app-electron} we provide an example of a system with
deterministic evolution by considering the case of a body with no
electric dipole moment and a magnetic dipole moment proportional to
spin.

We can rewrite our equations of motion in covariant form as
follows. In Fermi coordinates based on the worldline $\gamma$, the
time direction $(\partial/\partial t)^a$ on $\gamma$ coincides with
the unit tangent, $u^a$ to $\gamma$, so for example,
$F^{\textrm{ext}}_{i0}$ can be replaced by $u^a
F^{\textrm{ext}}_{ia}$. Similarly, spatial components of quantities on
$\gamma$ correspond to projections orthogonal to $u^a$, so, for
example, the spatial components $u^a F^{\textrm{ext}}_{ia}$ correspond
to the spacetime tensor $({\delta^c}_b +u^c u_b)u^a
F^{\textrm{ext}}_{ca} = u^a F^{\textrm{ext}}_{ba}$. The spatial derivative,
$\partial_i$, in Fermi coordinates can be expressed in terms of the
spatial projection of the covariant derivative $\nabla_\mu$ by using
the following formula for the Christoffel symbol in Fermi coordinates,
\begin{equation}\label{eq:covgamma}
\Gamma^{\alpha}_{\ \mu \nu}|_\gamma = a^\alpha u_\mu u_\nu
- u^\alpha u_\mu a_\nu - u^\alpha a_\mu u_\nu \,\, .
\end{equation}
Thus, $\partial_i F^{\textrm{ext}}_{\mu \nu}$ corresponds to
$({\delta^d}_c +u^d u_c)\nabla_d F^{\textrm{ext}}_{ab} + 2 a_c  u^e F_{e[b} u_{a]}$.
Finally, since, by
construction, the Fermi coordinate vector fields $\{(\partial/\partial
t)^a, (\partial/\partial x^i)^a\}$ are Fermi transported along
$\gamma$, it is clear that taking the time derivative, ``$d/dt_0$'',
of Fermi coordinate components along $\gamma$ corresponds to taking
the Fermi derivative of the corresponding tensors along $\gamma$. Here
the Fermi derivative, $D_F/d\tau$, is given by the usual covariant
derivative $D/d\tau \equiv u^a \nabla_a$ along $\gamma$
combined with a Lorentz boost in the plane defined by $u^a$ and $a^a$.
Specifically, for a scalar function $f$ on $\gamma$ we have
\begin{equation}
\frac{D_F}{d\tau} f = \frac{D}{d\tau} f \equiv u^a \nabla_a f ,
\end{equation}
for a covector field $f_a$ on $\gamma$ we have
\begin{equation}
\frac{D_F}{d\tau} f_a =\frac{D}{d\tau} f_a + 2 f^c a_{[a} u_{c]} ,
\end{equation}
and for a tensor field $f_{ab}$ on $\gamma$ we have
\begin{equation}
\frac{D_F}{d\tau} f_{ab} = \frac{D}{d\tau} f_{ab} + 2 f_a^{\ c}
a_{[b}u_{c]} + 2 f^c_{\ b} a_{[a}u_{c]}.
\end{equation}

In covariant notation, the evolution equations become
\begin{align}
m \delta a_a & = -a_a \delta m + \delta [ q F^{\textrm{ext}}_{ab} u^b ] +
a_a u_b u^c Q^{bd} F^{\textrm{ext}}_{cd} \nonumber \\
& \quad + \left( g_a^{ \ b} + u_a u^b \right) \bigg \{ \frac{2}{3} q^2
\frac{D_F}{d\tau}a_b - \frac{1}{2} Q^{cd}\nabla_b
F^{\textrm{ext}}_{cd} + \frac{D_F}{d\tau} \left( a^c S_{c b} - 2 u^d
Q^c_{\ [b} F^{\textrm{ext}}_{d]c} \right) \bigg \}
\label{eq:EOM-cov0} \\
\label{eq:spin-cov0}
\frac{D_F}{d\tau} S_{ab} & = 2 \left(g^c_{\ a} + u^c u_a\right) \left(g^d_{\ b} + u^d u_b\right) Q^e_{\ [c}F^{\textrm{ext}}_{d]e}\\
\label{eq:mass-cov0}
\frac{D_F}{d\tau} \delta m & = \frac{1}{2} Q^{ab} \frac{D_F}{d\tau}
F^{\textrm{ext}}_{ab}
+ \frac{D_F}{d\tau} \left(u_b u^c Q^{bd} F^{\textrm{ext}}_{cd} \right) \,\, .
\end{align}
In addition, we have $u^a S_{ab} = 0$ (see eq.~(\ref{spinCM}) above).  We remind the reader that these equations would have to be supplemented with a rule for the evolution of the electromagnetic dipole (arising from additional assumptions about the body) in order to comprise a ``closed,'' deterministic system.

By inspection, we see that eqs.~(\ref{eq:EOM-cov0}) and
(\ref{eq:mass-cov0}) simplify if we redefine the perturbed mass by
\begin{equation}
\delta m \rightarrow \delta \hat{m} \equiv
\delta m - u_b u^c Q^{bd} F^{\textrm{ext}}_{cd} \,\, .
\label{deltamhat}
\end{equation}
This corresponds to adding the standard expression, $-\vec{p} \cdot
\vec{E}{}^{\textrm{ext}}$, for electric dipole interaction energy
(with $\vec{E}{}^{\textrm{ext}}$ the external electric field in the
rest frame of the body) to the definition of perturbed mass. Note,
however, that we do not make any similar $-\vec{\mu} \cdot
\vec{B}{}^{\textrm{ext}}$ adjustment to the definition of perturbed
mass.  With this re-definition, eqs.~(\ref{eq:EOM-cov0}) and
(\ref{eq:mass-cov0})  become
\begin{equation}
\label{eq:EOM2}
m \delta a_a = -a_a \delta \hat{m} + \delta [q F^{\textrm{ext}}_{ab} u^b]
+ \left(g_a^{ \ b} + u_a u^b \right) \bigg \{ \frac{2}{3} q^2
\frac{D_F}{d\tau}a_b - \frac{1}{2} Q^{cd}\nabla_b
F^{\textrm{ext}}_{cd} + \frac{D_F}{d\tau} \left( a^c S_{c b} - 2 u^d
Q^c_{\ [b} F^{\textrm{ext}}_{d]c} \right) \bigg \}
\end{equation}
\begin{equation}\label{mass-evolution2}
\frac{D_F}{d\tau} \delta \hat{m} = \frac{1}{2} Q^{ab} \frac{D_F}{d\tau}
F^{\textrm{ext}}_{ab} \,\, .
\end{equation}

We finally rewrite eqs.~(\ref{eq:EOM2}), (\ref{eq:spin-cov0}) and (\ref{mass-evolution2})
in terms of the usual covariant
derivative $D/d\tau = u^a \nabla_a$ along $\gamma$, obtaining
\begin{align}
\label{eq:EOM-cov}
\delta [\hat{m} a_a ] & = \delta [ q F^{\textrm{ext}}_{ab} u^b ] + \left( g_a^{ \ b} + u_a u^b \right) \bigg \{ \frac{2}{3} q^2 \frac{D}{d\tau}a_b - \frac{1}{2} Q^{cd}\nabla_b F^{\textrm{ext}}_{cd} + \frac{D}{d\tau} \left( a^c S_{c b} - 2 u^d Q^c_{\ [b} F^{\textrm{ext}}_{d]c} \right) \bigg \} \\
\label{eq:spin-cov}
\frac{D}{d\tau} S_{ab} & = 2 \left(g_a^{\ c} + u_a u^c\right) \left(g_b^{\ d} + u_b u^d\right) Q^e_{\ [c}F^{\textrm{ext}}_{d]e} - 2 a^c S_{c[a}u_{b]} \\
\label{eq:mass-cov}
\frac{D}{d\tau} \delta \hat{m} & = \frac{1}{2} Q^{ab} \frac{D}{d\tau}
F^{\textrm{ext}}_{ab} + 2 Q_a^{\ b} F^{\textrm{ext}}_{bc} a^{[c}u^{a]}
\end{align}
where $\delta [\hat{m} a_a ] \equiv m \delta a_a + (\delta \hat{m}) a_a$.

The first term on the right side of equation \eqref{eq:EOM-cov} is
simply the corrected Lorentz force. The first term in curly brackets
is the Abraham-Lorentz-Dirac self force (see eq.~(\ref{eq:ALD-intro})
above).  The second term in curly brackets corresponds to the usual
dipole forces familiar from electrostatics and magnetostatics. The
final terms in curly brackets
are not usually considered in elementary treatments of
electromagnetic forces. We will show elsewhere \cite{us} that these terms
are responsible for producing behavior in an orbiting charged body
with spin and a magnetic dipole moment that is analogous to behavior
found for spinning black holes in binary orbits in general relativity.  The first term on the right side of eq.~\eqref{eq:spin-cov}
corresponds to the usual dipole torques familiar from electrostatics
and magnetostatics. The second term is the Thomas precession. Finally, we note that the (modified) perturbed mass $\delta \hat{m}$
is {\it not} constant in time.  However, in some cases---for example in the case discussed in appendix \ref{sec:app-electron}---there may be an alternative notion of mass that \textit{is} conserved.

Equations~(\ref{eq:EOM-cov})-(\ref{eq:mass-cov}) together with $u^a
S_{ab}=0$ provide the complete description of the first-order
deviation from Lorentz force motion, eq.(\ref{eq:lorentzforce}), that
can be derived from Maxwell's equations and conservation of
stress-energy alone, and comprise the main result of this paper.

\section{Self-consistent Motion}
\label{sec:self-consistent}

There are no mathematical difficulties that arise if one uses
eqs~(\ref{eq:EOM-cov})-(\ref{eq:mass-cov}) to obtain the lowest order
deviation from Lorentz force motion, eq.~(\ref{eq:lorentzforce}). The
lowest order quantities $m$, $q$, and $F^{\textrm{ext}}_{ab}$ are to
be viewed as ``given'', and the unperturbed worldline, $\gamma$, and
its associated quantities $u^a$, $a^a$, and $D a^a/d \tau$ are to be
viewed as having been determined by solving
eq.~(\ref{eq:lorentzforce}). We wish to determine the perturbed
worldline as well as the time evolution of the quantities $S_{ab}$,
$Q_{ab}$, $\delta q$, and $\delta \hat{m}$. The perturbed worldline is
described by a deviation vector, $X^a$, on the background worldline
$\gamma$, defined as follows: Consider the $\lambda$-dependent family
of curves, $\gamma(\lambda)$, introduced in subsection
\ref{subsec:parameters}. Parameterize each curve by proper time
$\tau$. In arbitrary fixed (i.e., $\lambda$-independent) coordinates $x^\mu$,
the family of curves is described by $Z^\mu(\lambda;\tau)$.  Then
the coordinate components of the deviation vector are given by $X^\mu
\equiv (\partial Z^\mu(\lambda;\tau)/\partial \lambda)|_{\lambda=0}$.
It follows that
\begin{equation}
\delta u^a = \frac{D}{d\tau} X^a
\label{deltau}
\end{equation}
and
\begin{equation}
\delta a^a = \frac{D}{d\tau} \delta u^a = \frac{D^2}{d\tau^2} X^a.
\label{deltaa}
\end{equation}
Furthermore, since $\delta F^{\textrm{ext}}_{ab} = \left. \left(
\frac{\partial}{\partial \lambda}\left(
F^{\textrm{ext}}_{ab}(\lambda)|_{\gamma(\lambda)} \right) \right)
\right|_{\lambda=0}$ we have
\begin{equation}
\delta F^{\textrm{ext}}_{ab} = X^c \nabla_c F^{\textrm{ext}}_{ab} +
F^{\textrm{ext}, (1)}_{ab},
\label{deltaF}
\end{equation}
where $F^{\textrm{ext}, (1)}_{ab}$ is the first order perturbation of
the external field arising in far-zone perturbation theory, evaluated
on $\gamma$. Again $F^{\textrm{ext}, (1)}_{ab}$ should be viewed as
``given'' (and, indeed, it would normally be assumed to vanish).  It
can then be seen explicitly that
eqs.~(\ref{eq:EOM-cov})-(\ref{eq:mass-cov}) together with
eq.~(\ref{deltau}) and $D \delta q/d \tau = 0$ comprise a system of
linear, first-order ordinary differential equations for $(X^a, \delta
u^a, S_{ab}, Q_{ab}, \delta q, \delta \hat{m})$.  If supplemented by
an appropriate evolution equation for $Q_{ab}$ these
equations have a unique solution for any given initial values of these
quantities. The initial $S_{ab}$ also must satisfy the center of mass
condition $S_{ab} u^a = 0$; this condition
on $S_{ab}$ is then preserved
by the evolution equation (\ref{eq:spin-cov}).  The solutions to these equations
are not plagued by the type of pathologies that occur for solutions to the Abraham-Lorentz-Dirac
equation (\ref{eq:ALD-intro}).

Nevertheless, the above equations do not give a fully satisfactory
description of the deviation from Lorentz force motion due to
self-force, spin, and electromagnetic dipole moment. Even if all of
the terms on the right side of eq.~(\ref{eq:EOM-cov}) are small compared to the Lorentz force at all times, the cumulative effects of these
terms over time should eventually make the deviation vector, $X^a$,
become large, at which point the description of motion as a linear
perturbation of a single, fixed Lorentz force trajectory cannot be
accurate.  The solution to this difficulty is not to go to higher
order in perturbation theory but to realize that although the
deviation from a single Lorentz force trajectory may eventually become
large at late times, the local deviation from {\it some} Lorentz force
trajectory should be small at all times.  As discussed in much more
depth in \cite{gralla-wald}, to implement the description of motion
as a locally small deviation from some (varying)
Lorentz force trajectory, we need to find a system
of ``self-consistent perturbative equations'' that satisfies the
following criteria: (1) They must have a well posed initial value
formulation. (2) They must have the same number of degrees of freedom
as the first order perturbative system, so that a correspondence can
be made between initial data for the self-consistent perturbative
equation and the first order perturbative system. (3) For
corresponding initial data, the solutions to the self-consistent
perturbative equation should be close to the corresponding solutions
of the first order perturbative system over the time interval for
which the first order perturbative description should be accurate.

An obvious first attempt to find such a self-consistent perturbative
equation would be to simply ``delete the $\delta$'s'' from
eqs~(\ref{eq:EOM-cov})-(\ref{eq:mass-cov}), i.e., take the corrected
equations of motion to be
\begin{align}
\label{eq:EOM-cov1}
\hat{m} a_a  & =  q F^{\textrm{ext}}_{ab} u^b  + \left( g_a^{ \ b} + u_a u^b \right) \bigg \{ \frac{2}{3} q^2 \frac{D}{d\tau}a_b - \frac{1}{2} Q^{cd}\nabla_b F^{\textrm{ext}}_{cd} + \frac{D}{d\tau} \left( a^c S_{c b} - 2 u^d Q^c_{\ [b} F^{\textrm{ext}}_{d]c} \right) \bigg \} \\
\label{eq:spin-cov1}
\frac{D}{d\tau} S_{ab} & = 2 \left(g_a^{\ c} + u_a u^c\right) \left(g_b^{\ d} + u_b u^d\right) Q^e_{\ [c}F^{\textrm{ext}}_{d]e} - 2 a^c S_{c[a}u_{b]} \\
\label{eq:mass-cov1}
\frac{D}{d\tau} \hat{m} & = \frac{1}{2} Q^{ab} \frac{D}{d\tau}
F^{\textrm{ext}}_{ab} + 2 Q_a^{\ b} F^{\textrm{ext}}_{bc} a^{[c}u^{a]}
\end{align}
together with $u^a S_{ab} = 0$.  In a sense, this is the simplest
modification we can make to the Lorentz force equation
(\ref{eq:lorentzforce}) that takes into account the first-order
perturbative effects of eqs.~(\ref{eq:EOM-cov})-(\ref{eq:mass-cov}).
However, it is easily seen that although
eqs.~(\ref{eq:EOM-cov1})-(\ref{eq:mass-cov1}) are ODEs and thus have
an initial value formulation, these equations violate criteria (2) and
(3) above. Specifically, the role of the acceleration, $a^a$, changes
from that of a zeroth order ``background'' quantity in
eq.~(\ref{eq:EOM-cov}) to that of an unknown dynamical variable in
eq.~(\ref{eq:EOM-cov1}).  Consequently, the terms involving
$Da^a/d\tau$ on the right side of eq.~(\ref{eq:EOM-cov1}) make this
equation be of higher differential order than the corresponding
perturbative equation (\ref{eq:EOM-cov}).  Thus, more initial data is
required for eqs~(\ref{eq:EOM-cov1})-(\ref{eq:mass-cov1}) than
for eqs~(\ref{eq:EOM-cov})-(\ref{eq:mass-cov}), in violation of
criterion (2). Criterion (3) also is violated.  Indeed, when $Q^{ab}$,
and $S_{ab}$ vanish, eq.~(\ref{eq:EOM-cov1}) reduces to the ALD
equation (\ref{eq:ALD-intro}), so even in this simple case,
eqs~(\ref{eq:EOM-cov1})-(\ref{eq:mass-cov1}) have solutions that
differ drastically from eqs~(\ref{eq:EOM-cov})-(\ref{eq:mass-cov}).
It is also worth noting that even when $q$ and $Q^{ab}$ vanish (so
there are no electromagnetic effects), if $S_{ab}$ is nonvanishing,
eq.~(\ref{eq:EOM-cov1}) is still of higher differential order than
eq.~(\ref{eq:EOM-cov}) on account of the $D (a^a S_{ab})/ d \tau$
term. The resulting unphysical degrees of freedom of the system
eqs~(\ref{eq:EOM-cov1})-(\ref{eq:mass-cov1}) together with $u^a S_{ab}
= 0$ give rise to solutions with ``helical motions'' in addition to
the expected solutions with inertial motion.

However, this difficulty can be overcome by the reduction of order
procedure described in the Introduction. We replace $D a^a/d \tau$ in
the two terms in which it appears on the right side of
eq.~(\ref{eq:EOM-cov1}) by the right side of eq.~(\ref{replaceadot}),
and we replace $a^a$ on the right sides of
eqs.~(\ref{eq:EOM-cov1})-(\ref{eq:mass-cov1}) by the right side of
eq.~(\ref{replacea}). The resulting system of equations---which we
will not write out explicitly here---still corresponds to Lorentz
force motion corrected by the perturbative effects of
eqs.~(\ref{eq:EOM-cov})-(\ref{eq:mass-cov}), but the resulting
equations now have highest derivative terms of exactly the same form
as eqs.~(\ref{eq:EOM-cov})-(\ref{eq:mass-cov}). Consequently,
criterion (2) holds, and we believe that criterion (3) also
holds. Thus, we believe that the reduced order form of
eqs~(\ref{eq:EOM-cov1})-(\ref{eq:mass-cov1}) supplemented by $u^a
S_{ab} = 0$ comprise a fully satisfactory system of self-consistent
perturbative equations that provide a description of motion that takes
into account the lowest order effects of self-force, spin, and
electromagnetic dipole moments. For comparison with what can be found
in textbooks on electromagnetism, we note that in the non-relativistic
limit\footnote{To obtain this equation, we drop all terms
in eq.~(\ref{eq:EOM-cov1}) that are quadratic or higher order in
velocity, $\vec{v}$, as well as terms linear in $\vec{v}$
that are multiplied by
$q^2$, $\vec{p}$, $\vec{\mu}$ or $\vec{S}$.}
and in ordinary vector notation, eq.~(\ref{eq:EOM-cov1}) takes the form
\begin{align}\label{eq:non-rel}
\vec{F} & = q\left(\vec{E} + \vec{v} \times \vec{B}\right)
  + \frac{2}{3}q^2\frac{d \vec{a}}{dt} +
  p_i \vec{\nabla}E^i + \mu_i \vec{\nabla}B^i  \nonumber \\
& \qquad + \frac{d}{dt} \left(
  \vec{S} \times \vec{a} - \vec{\mu} \times \vec{E} - \vec{p} \times
  \vec{B}\right) \,\, ,
\end{align}
where we have written $\vec{F} = m \vec{a}$ and it should again
be understood that the terms in $\vec{a}$ and $d \vec{a}/dt$ on the right
side of this equation should be eliminated by reduction of order.

Finally, as discussed in \cite{gralla-wald}, there is no reason to expect
self-consistent perturbative equations satisfying criteria (1)-(3) to
be unique. Although the reduced order form of
eqs~(\ref{eq:EOM-cov1})-(\ref{eq:mass-cov1}) supplemented by $u^a
S_{ab} = 0$ appears to be a fully satisfactory system
of self-consistent perturbative equations corresponding to the
corrections to Lorentz force given by
eqs~(\ref{eq:EOM-cov})-(\ref{eq:mass-cov}), other choices
of self-consistent perturbative equations are possible.  In particular, the
following system of equations arises naturally from the analysis of
the motion of extended bodies \cite{harte2009}: Define the ``force''
$f_a$ and ``torque''
$n_{ab} = n_{[ab]}$ in terms of the $4$-momentum $P^a$ and
and electromagnetic dipole $Q^{ab}$ by
\begin{equation}
  f_a = \frac{2}{3} (q/M)^3 h_{a}{}^{b} P^d (P^c \nabla_d
  F_{bc}^{\mathrm{ext}} - q g^{cf} F_{bc}^{\mathrm{ext}}
  F_{df}^{\mathrm{ext}} ) - \frac{1}{2} Q^{bc} \nabla_a
  F_{bc}^{\mathrm{ext}}
\label{f}
\end{equation}
\begin{equation}
  n_{ab} = 2 Q^{c}{}_{[a} F_{b]c}^{\mathrm{ext}} \,\, ,
\end{equation}
where
\begin{equation}
  M \equiv \sqrt{ - P_a P^a} ,
\end{equation}
and
\begin{equation}
  h_{ab} \equiv g_{ab} + P_a P_b /M^2 \,\, .
\end{equation}
(The first term in eq.~(\ref{f}) can be recognized as a reduced order
form of the Abraham-Lorentz-Dirac force.)
Then the tangent, $\dot{\gamma}^a$, to the center-of-mass worldline
(normalized so that $P_a \dot{\gamma}^a = -M$) is given in terms
of $P^a$, $Q^{ab}$ and the spin tensor $S^{ab}$ by
\begin{equation}
  M \dot{\gamma}_a = P_a - n_{ab} P^b/M - \frac{ S_{a}{}^{b} [q
      F_{bc}^{\mathrm{ext}} (P^c - n^{c}{}_{d} P^d/M) + M f_b ] }{ M^2
    - \frac{1}{2} q S^{cd} F_{cd}^{\mathrm{ext}} } \,\, .
\end{equation}
The evolution equations for $P^a$ and $S^{ab}$ are
\begin{equation}
  \dot{P}_a = q F_{ab}^{\mathrm{ext}} \dot{\gamma}^b + f_a
\end{equation}
\begin{equation}
  \dot{S}_{ab} = 2 P_{[a} \dot{\gamma}_{b]} + n_{ab} \,\, .
\label{Sevol}
\end{equation}
These equations are to be supplemented by $P_a S^{ab} = 0$, which, if
imposed as an initial condition, can be shown to be preserved under
evolution (see \cite{ehlers-rudolph}). At the
level of what is known purely from lowest order perturbation theory as
derived in section IV, there is no reason to prefer the system
(\ref{f})-(\ref{Sevol}) to the system
(\ref{eq:EOM-cov1})-(\ref{eq:mass-cov1}). However, in cases where
the self-force term in eq.~(\ref{f}) is negligible, the system
(\ref{f})-(\ref{Sevol}) yields an exact conservation
law when there is a Killing field that Lie derives
$F_{ab}^{\mathrm{ext}}$---a feature that would hold exactly for test body
motion. Therefore, it appears that in this respect
the system (\ref{f})-(\ref{Sevol})
is superior to the system (\ref{eq:EOM-cov1})-(\ref{eq:mass-cov1}).

\section{Summary}
We have given a rigorous and systematic treatment of particle motion in classical electromagnetism.  We considered a one-parameter-family of solutions to the Maxwell and matter equations (\ref{eq:max1})-(\ref{eq:Jcons}) and (\ref{eq:Tcons}) containing a body that ``shrinks down'' to zero size, mass, and charge according to the scaling assumptions of section \ref{sec:assumptions}.  We found that the lowest-order description of the body is that of a point particle \eqref{eq:J1-2} and \eqref{eq:T1form} moving according to the Lorentz force law \eqref{eq:lorentzforce}.  The first-order corrections to this motion---including self-force, dipole force, and spin force effects---are then given by our rigorous perturbative result (\ref{eq:EOM-cov})-(\ref{eq:mass-cov}).  Finally, we addressed the issue of finding a self-consistent perturbative equation associated with our perturbative result.  We argued that the naive self-consistent perturbative equations \eqref{eq:EOM-cov1}-\eqref{eq:mass-cov1} can be modified by ``reduction of order'' to provide appropriate self-consistent perturbative equations.  In the case of negligible spin and electromagnetic dipole moment, this reduces to the reduced-order ALD equation.

\bigskip

\noindent
{\bf Acknowledgements} This research was supported in part by NSF grants PHY04-56619 and PHY08-54807 to the University of Chicago.

\appendix

\section{Scaling Properties of the Self-field}
\label{sec:app-scaling}

This appendix derives the two
properties of $F_{\mu \nu}^{\textrm{self}}(\lambda, t, x^i)$ that were claimed
to hold at the
end of section \ref{sec:assumptions}
for families of current distributions satisfying
eq.~\eqref{eq:Jnew}.

We first show that the retarded field $F_{\mu
  \nu}^{\textrm{self}}(\lambda, t, x^i)$ satisfies the scaling relation,
eq.~(\ref{eq:Fform}). In global inertial coordinates
$(t, x^i)$, the Lorenz-gauge vector potential for the retarded solution
with source $J^\mu$ is given by
\begin{equation}
  A_\mu^{\mathrm{self}} (\lambda, t, x^i) = \int \mathrm{d}^3 x' \left[ \frac{ J_\mu( \lambda, t-|x^i-x'^i|, x'^j)} { |x^i-x'^i|} \right] \,\, ,
\label{AIntegrated0}
\end{equation}
where, in our case, we have $J^{\mu}(\lambda,t,x^i) = \lambda^{-2}
\tilde{J}^{\mu}(\lambda,t,[x^i-z^i(t)]/\lambda)$ with $\tilde{J}^\mu$ smooth
in all variables and of compact support in the spatial variables.
Now introduce scaled coordinates $\bar{x}^i = [x^i-z^i(t)]/\lambda$
and $\bar{x}'^i = [ x'^i - z^i(t) ]/\lambda$. Using these definitions
together with eq.~\eqref{eq:Jnew} and the coordinate transformation
$\bar{y}^i = \bar{x}'^i-\bar{x}^i$, it is easily seen that
\begin{equation}
  A_\mu^{\mathrm{self}} (\lambda, t, x^i) = \int \mathrm{d}^3 \bar{y}
  \left[ \frac{ \tilde{J}_\mu( \lambda, t-\lambda |\bar{y}^i|,
      \bar{x}^j + \bar{y}^j+ [z^j(t) - z^j(t-\lambda
        |\bar{y}^k|)]/\lambda) }{ |\bar{y}^i| } \right]
  . \label{AIntegrated}
\end{equation}
Since $z^i(t)$ is smooth, there
must exist a smooth function $V^i$ such that
\begin{equation}
  z^i(t) - z^i(t- \lambda |\bar{y}^j|) = \lambda |\bar{y}^j| V^i(t, \lambda |\bar{y}^j|) . \label{VDef}
\end{equation}
For later reference, we note
that since $x^i = z^i(t)$ defines a timelike curve,
it follows from the mean value theorem that  $|V^i| < 1$.
Substituting in eq.~(\ref{AIntegrated}), we see that the integrand is
smooth in $(\lambda, t, \bar{x}^i)$ and that the convergence
properties of the $\bar{y}$-integral are sufficient to allow us to
interchange differentiation with respect to $(\lambda, t, \bar{x}^i)$
with integration.
It follows that in global inertial coordinates,
$A_\mu^{\mathrm{self}}$ smooth in these variables, i.e., $A_\mu^{\mathrm{self}}$
is of the form
\begin{equation}
  A_\mu^{\mathrm{self}}(\lambda, t, x^i) = \tilde{A}_\mu ( \lambda , t , [x^i - z^i(t)]/\lambda) .
\end{equation}
where $\tilde{A}_\mu ( \lambda, t, \bar{x}^i)$ is smooth in all of its
arguments. Differentiating with respect to the spacetime variables, we
find that $F_{\mu \nu}^{\textrm{self}}(\lambda, t, x^i)$ takes the
form \eqref{eq:Fform} in global inertial coordinates. By the same type
of argument as given in sect. \ref{sec:assumptions}, it then follows
that $F_{\mu \nu}^{\textrm{self}}(\lambda, t, x^i)$ takes the general form
\eqref{eq:Fform} in arbitrary coordinates for which $\nabla_\mu t$ is timelike,
as we desired to show.

We now derive a certain falloff behavior for $F_{\mu
  \nu}^{\mathrm{self}}$ outside of the body, specifically that,
in terms of the parameters
\begin{equation}
  \alpha \equiv |x^i-z^i(t)|, \qquad \beta \equiv \lambda / \alpha , \label{AlphaBeta}
\end{equation}
we have
\begin{equation}
  \lambda F_{\mu \nu}^{\mathrm{self}} = \beta^2 \mathcal{F}_{\mu \nu}(t, \alpha, \beta, n^i )\label{eq:Fself-ab0}
\end{equation}
at least for $\alpha$ and $\beta$ in a sufficiently small neighborhood
of $(0,0)$, where $\mathcal{F}_{\mu \nu}$ is smooth in all of its
arguments and
\begin{equation}
n^i \equiv [x^i-z^i(t)]/\alpha \,\, ,
\end{equation}
so that $|n^i| = 1$.
An analogous statement applied to one-parameter families of metrics
was used in \cite{gralla-wald} in the analysis of gravitational
self-force; as explained further in \cite{gralla-wald}, this
corresponds to what is needed to justify the use of matched asymptotic
expansions. Matter distributions (if any) were not explicitly
considered in \cite{gralla-wald}, so the analogous behavior of metrics
in that reference was an assumption. Here, we prove that the
electromagnetic analog follows from a very natural form for the
source, namely eq.~\eqref{eq:Jnew}.

We start, again, with eq.~\eqref{AIntegrated0} for the Lorentz-gauge
vector potential $A_\mu^{\mathrm{self}} (\lambda, t, x^i)$ in global
inertial coordinates.  Initially, we assume that $\alpha > 0$, so that
$\beta \geq 0$ and $n^i$ are well defined.  We define
\begin{equation}
     w^i = x'^i - z^i(t- |x^j-x'^j|)  \label{wDef}
\end{equation}
and we use eq.~(\ref{VDef}) to write
\begin{equation}
z^i(t- |x^j-x'^j|) = z^i(t) - |x^j-x'^j| V^i(t, |x^j-x'^j|) \,\, .
\end{equation}
We also write
\begin{equation}
w^i = \lambda \bar{w}^i = \alpha \beta \bar{w}^i \,\, .
\label{wbarDef}
\end{equation}
The integrand in eq.~\eqref{AIntegrated0} is proportional to
$\tilde{J}_\mu(\lambda, t - |x^i-x'^i|, \bar{w}^j)$.  Since
$\tilde{J}_\mu$ is of compact support in the last argument, there
exists a positive constant $D$ such that only points satisfying
$|\bar{w}^i| < D$ contribute to the integral. Consequently, we may
assume that $|\bar{w}^i| < D$ in the following.

Equations (\ref{wDef})-(\ref{wbarDef}) together with the above definitions
of $\alpha$ and $n^i$ yield
\begin{equation}
x^i - x'^i = \alpha(n^i - \beta \bar{w}^i) + |x^j-x'^j| V^i(t,|x^k-x'^k|) \,\, .
\label{yalpha1}
\end{equation}
Thus, taking the squared magnitude of both sides of this equation, we obtain
\begin{equation}
 y^2 = | \alpha ( n^i - \beta \bar{w}^i) + y V^i(t,y)|^2.
\label{y2}
\end{equation}
where we have written
\begin{equation}
y \equiv |x^i-x'^i| \,\, .
\label{yDef}
\end{equation}
Treating eq.~(\ref{y2}) as though it were a quadratic equation in $y$
(even though $V^i$ depends on $y$), one finds that
\begin{equation}
  y = \frac{ \alpha |n^i - \beta \bar{w}^i | }{ 1 - |V^j|^2 }
  \left( |V^k| \cos \theta + \sqrt{1-|V^k|^2 +|V^k|^2 \cos^2 \theta} \right) \,\, ,
 \label{SepEq}
\end{equation}
where $\cos \theta$ is defined by
\begin{equation}
|V^k| \cos \theta \equiv \frac{V_i (n^i - \beta \bar{w}^i)}{|n^j- \beta \bar{w}^j|}
\label{cosDef}
\end{equation}
and $V^i$ is evaluated at $(t,y)$.
Since $y$ appears on both the left and
right sides, eq.~(\ref{SepEq}) is not a solution for $y$ but
rather is a relation that must be satisfied by the variables $y $, $t$,
$\alpha$, $\beta$, $n^i$, and $\bar{w}^i$.

As already noted below eq.~(\ref{VDef}), we have $|V^i| <
1$. Furthermore, since $|\bar{w}^i| < D$, if we restrict to $\beta <
1/D$, then $|n^i - \beta \bar{w}^i | > 0$. Consequently, for $\beta <
1/D$,
the right side of eq.~(\ref{SepEq}) is of the form of $\alpha$ times a
smooth, nonvanishing function of
$(t,y,n^i, \beta \bar{w}^i)$.
We
may therefore solve this equation for $\alpha$, obtaining a solution of the form
\begin{equation}
\alpha = y H(t,y,n^i,\beta \bar{w}^i)
\label{yalpha3}
\end{equation}
where $H$ is a smooth, nonvanishing function of all of its arguments.

Thus far, we have assumed that $\alpha > 0$, which implies $y > 0$
given our restrictions that $|\bar{w}^i| < D$ and $\beta <
1/D$. However, eq.~(\ref{SepEq}) makes perfectly good mathematical
sense---and $H$ in eq.~(\ref{yalpha3}) remains smooth---if we extend
the domain of $y$ and $\alpha$ to include $0$. (Indeed,
eq.~(\ref{SepEq}) makes sense---and $H$ remains smooth---even
for (unphysical) negative values of $y$, $\beta$, and $\alpha$.) We may
therefore apply the inverse function theorem at $y= \alpha = 0$ to conclude
that in a sufficiently small neighborhood of $\alpha=0$, $y$ can be
written as
\begin{equation}
 y = \alpha L(t, \alpha, n^i, \beta \bar{w}^j) \,\, ,
\end{equation}
where $L$ is a smooth, positive function of its arguments.

Rearranging \eqref{wDef}, we have
\begin{equation}
  x'^i = \alpha \beta \bar{w}^i + z^i(t-\alpha L(t, \alpha, n^j, \beta \bar{w}^k)) \,\, .
\label{xwbar}
\end{equation}
We now return to eq.~\eqref{AIntegrated0} and make the coordinate
transformation $x'^i \rightarrow \bar{w}^i$. Using eq.~(\ref{xwbar}),
we see that the Jacobian of this coordinate transformation takes the form
\begin{equation}
  \left| \frac{ \partial x'^i }{ \partial \bar{w}^j } \right| = (\alpha \beta)^3 W( t, \alpha, \beta, n^i, \bar{w}^j) \,\, ,
\end{equation}
where $W$ is a smooth function of its arguments.
Substitution into \eqref{AIntegrated0} finally shows that
\begin{equation}
    A_\mu^{\mathrm{self}}(\lambda, t, x^i) = \beta \int \mathrm{d}^3 \bar{w} \left( \frac{ \tilde{J}_\mu (\alpha \beta, t-\alpha L(t,\alpha, n^i, \beta \bar{w}^j) , \bar{w}^k ) }{ L(t, \alpha, n^i, \beta \bar{w}^j) } \right) W(t, \alpha, \beta, n^l, \bar{w}^h) . \label{AIntegrated2}
\end{equation}
The denominator here has been shown to be strictly positive over the range
of $\bar{w}^i$ for which
the numerator is non-vanishing. Both $L$ and $W$ are smooth and the integral
is being taken over a compact region. The
integral is therefore smooth in $t$, $\alpha$, $\beta$, and
$n^i$, i.e., there exists a smooth $\mathcal{A}_\mu (t,
\alpha, \beta, n^i)$ satisfying
\begin{equation}
  A_\mu^{\mathrm{self}} = \beta \mathcal{A}_\mu
\end{equation}
for sufficiently small $\alpha$ and $\beta$.
Differentiating to obtain the field strength shows that
\begin{equation}
  F_{\mu \nu}^{\mathrm{self}} = (\beta / \alpha) \mathcal{F}_{\mu \nu}(t, \alpha, \beta, n^i )\label{eq:Fself-ab}
\end{equation}
for some $\mathcal{F}_{\mu \nu}$ smooth in all of its arguments. Since
$\lambda = \alpha \beta$, it follows immediately that
$\lambda F_{\mu \nu}^{\mathrm{self}}$ satisfies eq.~(\ref{eq:Fself-ab0}).
We have derived eq.~(\ref{eq:Fself-ab0}) in global inertial coordinates,
but it is readily seen that this form is preserved under an arbitrary
smooth transformation to new coordinates for which $\nabla_\mu t$
is timelike.

\section{``Electron'' motion}\label{sec:app-electron}

It has been stressed in subsection \ref{subsec:motion} that the
perturbed
equations of motion that can be derived using only Maxwell's equations
and conservation of total stress-energy do not constrain the time
evolution of the electromagnetic dipole moment $Q_{ab}$. Thus, in
order to have a deterministic system, the equations we derived in
subsection \ref{subsec:motion} would have to be supplemented by an
evolution law for $Q_{ab}$ arising from the particular matter model
under consideration. A particularly simple and interesting case is to assume
that  for some (time independent) constant $C$ we have
\begin{equation}
  Q_{ab} = - C S_{ab} . \label{SpinDipole}
\end{equation}
From eqs.~\eqref{eq:DipoleDecomposition} and \eqref{eq:SpinVectToTens} together with
our center of mass condition $u^a S_{ab} = 0$, it is easily seen that
this is equivalent to having a vanishing
electric dipole moment of the body
and a magnetic moment  proportional to its spin, i.e.,
\begin{equation}
p_a = 0, \quad \quad \mu_a = C S_a \,\, .
\label{SpinDipole2}
\end{equation}
Since we have derived an evolution equation for $S_{ab}$,
eq.~\eqref{SpinDipole} serves to fix the time evolution of $Q_{ab}$ as
well. As discussed at the beginning of section
\ref{sec:self-consistent}, eqs~(\ref{eq:EOM-cov})-(\ref{eq:mass-cov})
together with eq.~\eqref{SpinDipole} and $u^a S_{ab} = 0$ comprise a
well-posed deterministic system for determining the perturbed motion. Note that
since the electric dipole moment vanishes, we have $\delta \hat{m}= \delta m$ (see
eq.~(\ref{deltamhat})).

Since many elementary particles satisfy eq.~(\ref{SpinDipole}), we may view the
self-consistent perturbative system  eqs.~(\ref{eq:EOM-cov1})-(\ref{eq:mass-cov1})
together with eq.~(\ref{SpinDipole}) as providing a simple, classical model for the motion of
an elementary particle, such as an electron. To write these equations  in a more familiar
form, it is useful to decompose $F_{ab}^{\mathrm{ext}}$ into
{\it rest-frame} electric and magnetic fields via
\begin{equation}
    F_{ab}^{\mathrm{ext}} = 2 u_{[a} E_{b]} + \epsilon_{abcd} u^c B^d . \label{eq:FieldDecomp}
\end{equation}
The evolution equation
\eqref{eq:spin-cov1}
for the angular momentum can then be written as
\begin{equation}
  \frac{\mathrm{D}}{\mathrm{d} \tau} S_a = - \epsilon_{abcd} u^b \mu^c B^d + 2 u_{[a} S_{b]} a^b .
\end{equation}
Since $\mu_a = C S_a$, the right side is orthogonal to $S_a$, so
\begin{equation}
  \frac{\mathrm{d}}{\mathrm{d} \tau} (S_a S^a) = 0 ,
\end{equation}
i.e., the spin vector does not change its magnitude under time evolution. Clearly, $\mu_a$ also does
not change its magnitude, i.e., the body has a ``permanent'' magnetic dipole moment.

The evolution equation \eqref{eq:mass-cov1} for mass reduces to
\begin{equation}
\label{mvar}
  \frac{\mathrm{d}}{\mathrm{d} \tau} \left(m + \mu_a B^a \right) =0 \,\, ,
\end{equation}
where we have dropped the ``hat'' on $m$ since there is no distinction between
$\hat{m}$ and $m$ when the electric dipole moment vanishes.
Although $m$ does not remain constant, it is clear that we may
define a new quantity
\begin{equation}
  m_{*} \equiv m + \mu_a B^a .
  \label{mStar}
\end{equation}
that is conserved. Note that the standard expression for interaction
energy, $- \mu_a B^a$, is being \textit{subtracted} from $m$ in this
equation---rather than added to $m$---to define $m_{*}$.

Note that the failure of the ``rest mass'' $m$ to be constant resolves a paradox concerning what
one is taught in elementary physics courses: On one hand, one is (correctly) taught that an external
magnetic field can ``do no work'' on a body, so a body moving in an external magnetic field cannot
gain energy. On the other hand, one is (also correctly) taught that a magnetic dipole released in a non-uniform external magnetic field will gain kinetic energy. Where does this kinetic energy come from?
Equation (\ref{mvar}) shows that it comes from the rest mass of the body.\footnote{By contrast, for a ``permanent electric dipole'', i.e., $Q_{ab} = 2 u_{[a}p_{b]}$ with $D_F Q_{ab} / d\tau = 0$, it follows from eq.\eqref{eq:mass-cov1} that the \textit{original} mass $m = \hat{m} + p_a E^a$ is conserved.  The kinetic energy gained by an electric dipole released in a non-uniform electric field comes from the work done on the body by the electric field.}

We now consider the motion of the body, as described by eq.~\eqref{eq:EOM-cov1}.
Using eq.~(\ref{SpinDipole}), we obtain
\begin{eqnarray}
  m a_a = q E_a + (\delta_{a}{}^{b} + u_a u^b) \left( \frac{2}{3} q^2 \frac{\mathrm{D} a_b}{\mathrm{d} \tau} + \frac{1}{2} \epsilon^{cdef} u_c \mu_d \nabla_b F_{ef}^{\mathrm{ext}}  \right) \nonumber
  \\
  ~ + \epsilon_{abcd} u^b S^c \frac{\mathrm{D}}{\mathrm{d} \tau} ( C E^d - a^d ) + 2 \mu_{[a} B_{b]} (C E^b - a^b) ,
  \label{ElectronMotion}
\end{eqnarray}
where the terms involving acceleration on the right side are understood
to be eliminated by reduction of order.\footnote{In particular, a term proportional to  $\epsilon_{bcdf} a^c \mu^d E^f$ has been set to zero since  $a^a$ is to be replaced by $qE^a/m$.}
If $q \neq 0$, it is conventional to write $C$ in terms of a dimensionless $g$-factor
defined by $C = qg/2m$, in which case
\begin{equation}
  C E^a - a^a = \frac{(g-2) q}{2 m} E^a \,\, ,
\end{equation}
where we have again used $a^a = qE^a/m$ for terms on the right side.
In the case of an electron or muon, $g$ is very nearly $2$, so the last two terms on the right
side of eq.~\eqref{ElectronMotion} should be negligible. However, $g$ can be very different from $2$ for composite particles, in which case the last two terms in eq.~\eqref{ElectronMotion} need not be negligible.

Another point worth noting with regard to both  eq.~\eqref{eq:EOM-cov1} and eq.~\eqref{ElectronMotion} is that the Lorentz-Dirac self-force is not the only term involving $\mathrm{D} a_b/ \mathrm{d} \tau$. A ``jerk'' force also occurs in conjunction with the spin of the body in the second line of
eq.~\eqref{ElectronMotion}. (Note that this force is orthogonal to the ALD force.) For a body of charge $q$ equal to the charge of the electron $e$, and with spin of order $\hbar$, the ratio of the magnitude of the
``spin jerk'' term to the
ALD term is of order $(g-2) \hbar/e^2 = (g-2)/\alpha$. Thus, if $g \neq 2$, the ``spin jerk'' term can dominate over the ALD term. However, the effects of the ``spin jerk'' term probably do not accumulate significantly over time.

It also interesting to ask when the ALD self-force becomes comparable to the Lorentz force.
In this situation, it is likely that higher order corrections to the self-force will be large, and our
perturbative equations are unlikely to be reliable. It is easily seen that for an external electromagnetic
field that oscillates with frequency $\omega$, the condition that the ALD self-force is much smaller
than the Lorentz force is simply
\begin{equation}
q^2 \omega \ll m \,\, .
\end{equation}
For an electron, this reduces to
\begin{equation}
\alpha  \hbar \omega \ll m \,\,  .
\end{equation}
It seems unlikely that a classical description of electron motion would be possible
in any case if this condition were violated.

Another criterion for the validity of our perturbative equations is that the change in mass under
time evolution---which is a perturbative effect---be small compared with the mass, i.e., that
\begin{equation}
|\mu| |B| \ll m
\end{equation}
If $g \sim 1$ and $|q| = e$, this reduces to
\begin{equation}
|B| \ll \frac{m^2}{ e \hbar} \sim \left( 4 \times 10^{13} \, \mathrm{G} \right) \left( \frac{ m}{ 500 \, \mathrm{keV} } \right)^2
\end{equation}
Magnetic fields expected to exist near some neutron stars are expected to violate this bound. It seems
unlikely that any presently known classical equations of motion can adequately describe the behavior
of electrons when this bound is violated.

\end{document}